\documentstyle[aps,prb,epsf,twocolumn]{revtex}

\newcommand{\vev}[1]{\left\langle #1 \right\rangle} 

\newcommand{\bpm}[1]{\left(\matrix{#1}\right)}
\newcommand{\epm}{\end{array}\right)}
\newcommand{\bmm}{\begin{matrix}}
\newcommand{\emm}{\end{matrix}}
\renewcommand{\v}[1]{{\bf #1}} 
\newcommand{\Latt}{\hbox{Latt}}
\renewcommand{\t}[1]{{\tilde #1}}
\newcommand{\cK}{ {\cal K} } 
\newcommand{\cL}{ {\cal L} } 

\newcommand{\etal} {{\it et al.}}
\newcommand{\lookherea}{16}
\newcommand{\lookhereb}{11}

\begin{document}
\draft
\wideabs{
\title{Critical points in edge tunneling between
generic FQH states}
\author{Joel E. Moore$^a$ and Xiao-Gang Wen$^{b,}$\cite{wenweb}}
\address{$^a$ Bell Labs Lucent Technologies, 600 Mountain Avenue,
Murray Hill, NJ 07974 \\
$^b$Department of Physics and Center for Materials Science
and Engineering,\\
Massachusetts Institute of Technology,
Cambridge, MA 02139}
\date{\today}
\maketitle
\begin{abstract}
A general description of weak and strong tunneling fixed points is
developed in the chiral-Luttinger-liquid model of quantum Hall edge
states.  Tunneling fixed points are a subset of ``termination'' fixed
points, which describe boundary conditions on a multicomponent edge.
The requirement of unitary time evolution at the boundary gives a
nontrivial consistency condition for possible low-energy boundary
conditions.  The effect of interactions and random hopping on fixed
points is studied through a perturbative RG approach which generalizes
the Giamarchi-Schulz RG for disordered Luttinger liquids to broken
left-right symmetry and multiple modes.  The allowed termination
points of a multicomponent edge are classified by a $B$-matrix with
rational matrix elements.  We apply our approach to a number of
examples, such as tunneling between a quantum Hall edge and a
superconductor and tunneling between two quantum Hall edges in the
presence of interactions.   Interactions are shown to induce a continuous
renormalization of effective tunneling charge for the integrable case
of tunneling between two Laughlin states.  The correlation functions
of electronlike operators across a junction are found from the $B$
matrix using a simple image-charge description, along with the induced
lattice of boundary operators.  Many of the results obtained are
also relevant to ordinary Luttinger liquids.
\end{abstract}

}

\section{introduction}
Edge states of quantum Hall liquids have attracted continuous
attention since their importance was pointed out by
Halperin.~\cite{halperin} The only gapless excitations of a
two-dimensional electron gas on a quantum Hall plateau are the edge
excitations, since the bulk is an incompressible quantum
liquid.~\cite{laughlin} A quantum Hall (QH) state with a single
condensate, such as $\nu = 1$ or $\nu = 1/3$, has a single bosonic
mode of edge excitations, which can be thought of as hydrodynamic
perturbations of the Hall droplet.~\cite{wen1,wenrev0} For these
states all excitations propagate in a single direction (the edge is
`chiral'), the direction of the classical $E \times B$ drift at the
edge due to the confining field $E$ producing the edge.  The
chiral-Luttinger-liquid ($\chi$LL) model of edge structure predicts
that more complicated states, such as the integer QHE states with $\nu
\geq 2$, have multiple modes of edge excitations with generally
different velocities of propagation, and that some FQHE states have
modes traveling in both directions along the edge, as verified for the
$\nu = 2/3$ edge by numerical
calculations.~\cite{johnson,wenrev1,moore1}

This article studies the possible ways of tunneling electrons or
quasiparticles between one FQHE edge and another or between an FQHE
edge and a metallic or superconducting contact.  Such tunneling is of
interest because, at weak coupling (junction conductance much less
than $\frac{e^2}{h}$), tunneling experiments provide the most
sensitive probe available of edge properties; at strong coupling
(junction conductance of order $\frac{e^2}{h}$) tunneling between
simple edges is one of the few examples of a solvable, experimentally
accessible nonequilibrium interacting system.~\cite{fendley} In the
remainder of the introduction we outline our results and then briefly
review the comparison of the chiral-Luttinger-liquid description to
existing tunneling experiments.

The first part of this article sets up a framework to describe
possible tunneling fixed points in chiral Luttinger liquids and
applies it to a number of examples.  The best-known example of this
type of tunneling is between two one-component edges such as $\nu =
1/3$.  Quasiparticle tunneling at finite temperature across a slight
constriction in a single $\nu = 1/3$ quantum Hall bar becomes stronger
and stronger as the temperature is lowered (quasiparticle tunneling is
``relevant'' in renormalization-group language) until the constriction
becomes large and the system can be described as weak electron
tunneling between two separated $\nu = 1/3$ edges.  The crossover
between these two fixed points can also be driven by applying a
voltage across the junction at zero temperature.  The fascinating,
experimentally accessible physics of tunneling between quantum Hall
edges motivates a general study of what fixed points are possible in
this system.

Our approach starts by considering the different ways in which two
edges can be joined, or one edge can be terminated.  The class of
bosonic boundary conditions we consider is not formally complete (the
set of all allowed conformal boundary conditions for even a single
boson is unknown) but includes all previously treated cases plus many
others.  The requirement of unitary time evolution gives a physical
restriction on possible tunneling junctions or edge terminations.  As
an example of the results, it is possible for a $\nu = 1$ quantum Hall
edge to join smoothly to a $\nu=1/9$ edge in the presence of a
superconductor (acting as a charge reservoir), but not to a $\nu =
1/3$ edge: joining a $\nu=1$ and $\nu=1/3$ edge requires an entropy
flow to the surroundings.  The experimentally relevant correlation
functions for electronlike operators across the junction can be
calculated from this framework, and have a simple ``image-charge''
description when the tunneling problem is folded onto the half-line.
We present a solvable model in which interactions between edges give a
continuous variation with interaction strength of effective tunneling
charge and the $I-V$ curve.  This suggests that once interactions
between edges are considered, nonuniversal features can appear in
tunneling properties.

The last section develops a perturbative RG analysis of the effect on
the edge interaction matrix $V$, defined below, of random hopping of
quasiparticles between edge modes.  The tunneling properties of an
edge are affected by $V$ for both weak and strong tunneling, and for
weak tunneling such random hopping (introduced to model impurity
scattering) is required to obtain universal
behavior~\cite{kfp,kane,moorewen}.  The agreement discussed below
between the chiral-Luttinger-liquid and composite-fermion approaches
requires such hopping to drive the edge from a nonuniversal starting
point to a universal strong-coupling fixed point.  The importance of
$V$ for tunneling motivates a study of what form $V$ should take in
real systems, where some degree of impurity scattering is always
present.

We find that in all principal hierarchy states, random hopping of
quasiparticles drives the $V$ matrix to fixed points where charge and
neutral excitations are decoupled.  Without random hopping, in general
all eigenmodes carry charge.  When the neutral modes have the same
velocity, as in IQHE and main-sequence FQHE $\nu = n/(2n \pm 1)$
edges, each edge has one fixed point and these fixed points are
exactly solvable, as shown by Kane, Fisher, and Polchinski
(KFP),~\cite{kfp,kane} and have higher symmetry than the generic clean
system (``symmetry restoration by disorder'').  Edges with neutral
modes in both directions such as $\nu = 5/7$ can have several
charge-neutral separated fixed points, some of the solvable type found
by KFP and some unsolvable at present.  The RG flows have a simple
description in terms of the ``boost'' and rotation coordinates for
multicomponent $\chi$LLs.  The RG flow equations can also be applied
to the formation of the ``chiral metal'' phase~\cite{chalker,balents}
and other coupled Luttinger liquid problems.

\begin{figure}
\epsfxsize=3.75truein
\centerline{\epsffile{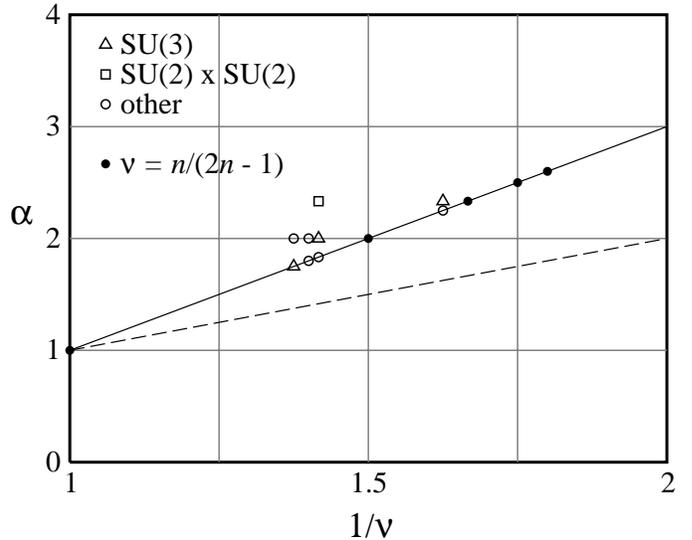}}
\caption{The tunneling exponent $I \propto V^{\alpha}$ for quantum Hall
states in the range $1/2 < \nu < 1.$  The solid line is the prediction
of the compressible-state theory of
Shytov, Levitov, and Halperin$^{\lookherea}$ for infinitely many channels.
The dotted line is $\alpha = 1/\nu$.  Solid circles are the main-sequence
edges with one phase per edge, and other shapes describe fixed points of
various symmetry classes in edges with multiple phases.$^{\lookhereb}$
The states shown are all principal hierarchy states up to 4th level:
the main-sequence states
plus $\nu = 8/11,5/7,12/17,8/13$.}
\label{figone}
\end{figure}

We review one class of experiments to illustrate the current status of
different descriptions of tunneling into edge states.  The tunneling
$I$-$V$ exponent from a Fermi liquid to an edge state is a sensitive
measurement of edge structure~\cite{milliken} (Fig.~\ref{figone})
which has received much attention since the results of Grayson {\it et
al.}~\cite{grayson} showing deviations from the predicted curve for
filling fractions $1/3 < \nu < 1$.  Other measurements~\cite{chang} do
show a broad plateau in rough agreement with theory, although the
plateau is displaced from its expected location, perhaps because the
density profile at the edge is not simply a sudden
decline.~\cite{levitov}

In general, an edge can have multiple edge phases, leading to different
tunneling exponents. Thus we can use tunneling to study transitions between
those edge phases.
Note that for each edge with multiple phases, exactly one phase has an
electron tunneling exponent lies
on the curve predicted by the composite-fermion (CF) approach of
Shytov, Levitov, and Halperin~\cite{shytov,levitov} in the limit of an
infinite number of channels.  Assuming this phase is the stable one at
finite temperature, the $\chi$LL and CF approaches appear to include
the same essential physics, even though the former assumes an
incompressible bulk state and the latter a compressible bulk state.
The currently unsolvable fixed points are relevant for experiments
(e.g., the phase on the CF line at $\nu = 5/7$) and are distinguished
by having relevant disorder operators coupling neutral modes in
opposite directions.  This suggests that the solvability of some fixed
points is a consequence of their chiral nature.  It is worth pointing
out that Fig.~\ref{figone} shows that smoothly varying physical
properties (in this case, the tunneling exponent) can emerge even
though the number of edge modes changes discretely.

The outline of the paper is as follows.  Section II sets up a
framework for describing the fixed points of a general quantum Hall
edge with a junction or terminus.  The description can be carried out
without reference to the details of the microscopic action except for
two essential formulas, which we derive in section V.  Section III
applies the framework to a number of cases of interest, obtaining the
critical points for a quantum Hall edge coupled to a superconductor
and two coupled quantum Hall edges.  Section IV studies the effects of
interactions on tunneling in a solvable model, where interactions lead
to a continuous change in the ``effective'' (i.e., observed in
tunneling experiments) charge and filling fraction.  Section V derives
the correlation functions and selection rules for boundary operators
in a chiral Luttinger liquid, which can be understood from an
``image-charge'' picture as in electrostatics.  Section VI examines
the effect of random hopping on the interaction matrix of a chiral
Luttinger liquid, including charge-neutral separation at the edge,
equilibration of velocities, and the basin of attraction of the fixed
points of Fig. \ref{figone}.  Section VII contains a brief summary of
our main conclusions.

\section{General description of critical points}
\subsection{Characterization of an Abelian edge}

In the following we will give a general description of edges of Abelian
quantum Hall states, emphasizing the data needed to characterize such
edges.  An Abelian edge state with $k$ branches is described by $k$ bosonic
fields $\phi_a$, $a=1,..,k$. The edge is characterized by a
$k$-dimensional symmetric real matrix $K$, a $k$-component charge
vector $\v q$, a lattice $\Gamma_c$ that determines the allowed charge
excitations, and a $k$-dimensional symmetric real matrix $\Delta$ that
describes the interaction between edge branches.  The total
electric charge density $\rho_e$ and current $j_e$ can be expressed
through ${\v \phi_a}$ and ${\v q}$:
\begin{eqnarray}
 \rho_e &=& q_a \rho_a, \ \ \ \rho_a =\frac{1}{2\pi}\partial_x \phi_a
\nonumber\\
 j_e &=& q_a j_a, \ \ \ j_a =-\frac{1}{2\pi}\partial_t \phi_a.
\end{eqnarray}
The allowed charge
excitations are created by the vertex operators
\begin{equation}
V_{\v n} = e^{i\v n\cdot {\v \phi}}, \ \ \v n \in \Gamma_c.
\end{equation}
Some $V_{\v n}$ add charge to the edge and others transfer electrons
or quasiparticles between different edge branches but are neutral
overall. The charge and the statistics of $V_{\v n}$ are given by
\begin{equation}
 Q = \v n^T  K^{-1} \v q, \ \ \theta =\pi \v n^T  K^{-1} \v n.
\end{equation}
The vectors in $\Gamma_c$ satisfy
\begin{eqnarray}
\v n^T  K^{-1} \v n' & =& \hbox{integer}, 
\nonumber\\
\v n, \v n' \in \Gamma_c.
\end{eqnarray}
Thus the charge excitations in $\Gamma_c$ are bosons or fermions with
trivial mutual statistics.  These vertex operators can appear in the
Hamiltonian if they have bosonic statistics. In many cases, the
lattice $\Gamma_c$ is generated by electron operators, so we will
refer to $\Gamma_c$ as the E-lattice (although $\Gamma_c$ sometimes
contains operators that transfer quasiparticles between different
branches).

The quasiparticle operators are also labeled by points in a lattice,
$\Gamma_q$:
\begin{equation}
\Gamma_q = \{ \v n|\v m^T K^{-1} \v n = \hbox{integer, for all } \v m \in
\Gamma_c \}
\end{equation}
We will call this lattice the quasiparticle lattice or Q-lattice.
Note that the Q-lattice $\Gamma_q$ is the dual lattice of the
E-lattice $\Gamma_c$. From the definition of the E-lattice, we see
that $\Gamma_c \subset \Gamma_q$.
Since we are going to discuss many lattices in this paper, we find it is
convenient to use a matrix to describe a lattice.
We will say a lattice $\Gamma$ is described by a matrix $M$
if the column vectors of the matrix generate the lattice.
We will denote such a lattice as $\Gamma = \Latt(M)$. Also we will use
$W\Gamma$  to denote the transformed lattice of $\Gamma$ by $W$:
$W\Gamma \equiv \Latt(WM)$.
Under this notation, we can write the E-lattice
$\Gamma_{c} = \Latt(C)$, where
the $k$ by $k$ matrix $C$ satisfies $C^T K^{-1} C=$ integral matrix.
The Q-lattice can be found to be
\begin{equation}
\Gamma_q = \Latt(K (C^T)^{-1}).
\end{equation}

The scaling dimensions of all vertex operators (electron or
quasiparticle operators) $V_{\v n}$
can be determined
from a single matrix $\Delta$, which depends on $ K$ and
interaction strengths between edge branches:
\begin{equation}
 h(\v n) = \frac12 \v n^T  \Delta\v n.
\end{equation}
Our reason for emphasizing $\Delta$ here rather the
``velocity'' matrix $V$ appearing in the $\chi$LL action
\begin{equation}
\label{action}
S_0 = {1 \over 4 \pi} \int{dx \, dt\, [K_{ij} \partial_x \phi_i
\partial_t \phi_j +
V_{ij} \partial_x \phi_i \partial_x \phi_j]}
\end{equation}
is that
the universal properties of the junction depend on scaling dimensions,
determined by $\Delta$; $V$ contains additional information (e.g. the
velocities of the eigenmodes) which is unnecessary for this
section.~\cite{moorewen}
For example, the (equal-space) correlation of $V_{\v n}$ has a form
\begin{equation}
 \vev{V_{\v n}(x, t) V^\dag_{\v n}(x,0)} \sim t^{-2h(\v n)}
\end{equation}

The matrix $\Delta$ is not an arbitrary real symmetric matrix.
If we introduce $K_{1/2}$ through:
\begin{equation}
K = K_{1/2}^T \left(\matrix{I_{k_+} & 0  \cr
          0   &-I_{k_-}}\right)  K_{1/2},
\end{equation}
then $\Delta$ can be expressed as
\begin{equation}
\Delta =  K_{1/2}^{-1}  B_{st}^{-2} ( K_{1/2}^T)^{-1}
\label{boost}
\end{equation}
where $B_{st}$ is the boost matrix introduced in \cite{moorewen} and
reviewed in section VI:
\begin{equation}
 B_{st} = \exp
\bpm{0 & b \cr
b^T& 0}
\end{equation}
and $b$ is a real $k_+\times k_-$ matrix.

\subsection{Termination of an Abelian edge}

First, let us consider the general problem of termination of an
Abelian edge at a point.  We know that a 1D electron gas contains two
branches, one right-moving and one left-moving. Such a 1D system can
be terminated at a point (say $x=0$) and we can have a system on a
half-line $0< x < \infty$. In contrast, the edge state of a $\nu=1/m$ QH
state contains only one right-moving branch. Such an edge state cannot
be terminated at any point without violating unitarity (alternately,
the Hamiltonian would be non-Hermitian).  Now the question is
when a generic Abelian edge described by $(\t K,\t{\v q}, \t
\Gamma_c,\t\Delta)$ can be terminated at a point, and if the edge
$(\t K,\t{\v q}, \t \Gamma_c,\t\Delta)$ can be terminated, how to
characterize the different ways in which the edge terminates.  Our
motivation for considering termination of an edge is that a tunneling
junction in an edge can be considered as a special kind of termination,
as in the next subsection.

We find that the edge $(\t K,\t{\v q}, \t \Gamma_c,\t\Delta)$ can
be terminated consistently (section V)
if there is a $2k$ by $k$ matrix $B$
that satisfies
\begin{eqnarray}
&&  B^T\t K^{-1} B = 0, \nonumber\\
&& \det(B^TB) \neq 0 \nonumber\\
&& \Gamma_B \equiv \Latt(B) \subset \t \Gamma_{c0}
\label{topos}
\end{eqnarray}
where $k=\dim(\t K)/2$ and
\begin{equation}
 \t \Gamma_{c0} =\{\v n| \v n \in \t \Gamma_c, \v n^T \t K^{-1} \v n
=\hbox{even}\}
\end{equation}
(i.e., the points in $\t \Gamma_{c0}$ describe the bosonic vertex operators).
The physical meaning of the first two conditions is that there are $k$
vectors of length $2k$ (the columns of the matrix $B$) which are null
in the indefinite quadratic form $K^{-1}$, orthogonal in $K^{-1}$,
and linearly independent.  Previously Haldane~\cite{haldane}
described charge-neutral
null vectors of $K^{-1}$ as ``topological instabilities,'' which allow
oppositely directed edge modes to localize each other and drop out of
the low-energy theory; thus condition (\ref{topos}) is that there
be $k$ independent topological instabilities orthogonal in $K^{-1}$.
(We have relaxed the condition of charge-neutrality in order to include
situations involving coupling to a superconductor.)
In order for $B$ to exist, $\t K$ must have the same number of
positive and negative eigenvalues (the edge has the same number of
right-moving and left-moving branches).

If more than one $B$ exists, then the edge
$(\t K,\t{\v q}, \t \Gamma_c,\t \Delta)$ can be
terminated in more than one way.
In other words, the boundary at $x=0$ can have more than one fixed point.
These different fixed points will be referred as different terminations
of the edge.

For a termination labeled by the matrix $B$, the fields $\t {\v \phi}$
satisfy the following boundary conditions at the termination point
\begin{equation}
 B^T \t {\v \phi} = 2\pi \v n,\ \ \ \v n = \hbox{integral vectors.}
\end{equation}
Note that $\v \phi$ can satisfy different boundary conditions 
(labeled by $\v n$) even for a
single type of terminate labeled by $B$.

The allowed charge excitations (vertex operators) at the boundary are
labeled by points in a $k$-dimensional
lattice $\Gamma_{q_B}$ (called the boundary Q-lattice):
\begin{eqnarray}
&& \Gamma_{q_B} = \\
&&\Latt\left(
\t K B(B^TB)^{-1} -\frac12 B(B^TB)^{-1}B^T\t KB(B^TB)^{-1}
\right)
\label{GaqB}
\nonumber
\end{eqnarray}
The boundary quasiparticle operators have a form
$V^b_{\v l} = e^{i\v l^T \t{\v \phi}}$, $\v l \in \Gamma_{q_B}$.
%
The scaling dimension of $V^b_{\v l}$ is given by
\begin{equation}
h^b(\v l) = \v l^T  \t K^{-1}B (B^T \t\Delta B)^{-1} B^T \t K^{-1} \v l
\end{equation}
This is one of the main results of this paper.

In the above discussion of termination points, we have ignored any symmetry
properties and the related selection rules. In particular, the boundary
condition characterized by $B$ may not conserve electric charge.
As a result, a boundary vertex operator may not carry a definite electric
charge.
In order for the termination labeled by $B$ to conserve the electric
charge, we must require the $B$ matrix to satisfy
\begin{equation}
 B^T \t K^{-1} \t {\v q} =0
\label{BCh}
\end{equation}
For the charge conserving termination points, the electric charge of a
boundary vertex operator $V^b_{\v l}$ is found to be
\begin{equation}
 Q=\t {\v q}^T \t K^{-1} \v l
\end{equation}

For a general termination described by $B$, there are $k$ combined charges
that are conserved near the boundary. Their densities are given by
\begin{equation}
\rho_a = B_{ba} \partial_x \t\phi_a/2\pi
\end{equation}
The boundary operator $V^b_{\v l}$ carries definite values of these $k$
combined charges:
\begin{equation}
 \v Q =  B^T \t K^{-1} \v l
\end{equation}

\subsection{Tunneling junction in an Abelian edge}

Now let us consider a more practical problem -- a tunneling junction in an
Abelian edge. Assume the edge is described by $(K,\v q, \Gamma_c, \Delta)$.
The tunneling is described by vertex operators in the Hamiltonian,
$V_{\v n}$, $\v n \in \Gamma_c$ (the lattice $\Gamma_c$ determines which
vertex operators are allowed).
If the charge is conserved, only the neutral vertex operators in
$\Gamma_c$ can
be added to the Hamiltonian to describe tunneling. These neutral vertex
operators describe the different types of electron/quasiparticle
tunneling between different edge branches. We would like to remark that the
tunneling junction referred to here can also be viewed as a defect or an
impurity on the edge.

To describe the possible quantum fixed points of the tunneling junction,
we can fold the edge in $(-\infty, 0)$ on top of the edge in $(0,\infty)$ by
introducing $2k$ fields $\t \phi_a$ on $(0,\infty)$:
\begin{eqnarray}
 \t \phi_a (x) &=& \phi_a(x) \nonumber\\
 \t \phi_{k+a} (x) &=& - \phi_a(- x) \nonumber\\
 x > 0 && a = 1,..,k.
\end{eqnarray}
The resulting edge is described by
$(\t K,\t {\v q}, \t \Gamma_c, \t \Delta)$:
\begin{eqnarray}
\t K &=& \left(\matrix{K&0 \cr 0& -K}\right) \nonumber\\
 \t \Delta &=& \left(\matrix{\Delta&0\cr0&\Delta} \right)\nonumber\\
 \t \Gamma_c &=& \Gamma_c \oplus \Gamma_c \nonumber\\
 \t {\v q} &=& \left(\matrix{{\v q}\cr {\v q}} \right).
\end{eqnarray}
The edge is terminated at $x=0$. Now the problem of the different
fixed points of a tunneling junction becomes a problem of different ways that
the edge $(\t K,\t {\v q}, \t \Gamma_c, \t \Delta)$ can terminate at $x=0$.
More precisely, each fixed point of the tunneling junction correspond to
a way in which the edge $(\t K,\t {\v q}, \t \Gamma_c, \t \Delta)$
terminates.

\subsection{Joining of two Abelian edges}

Next, let us consider the problem of possible different edge states
for a given QH liquid. There is the possibility that different edge
potentials and/or electron interactions can lead to different edge
states without changing the bulk QH liquid.  Indeed, it has been shown
that at zero temperature there are multiple stable edge phases of some
disordered FQH states with neutral modes in both
directions~\cite{moorewen}, such as $\nu=5/7$.  If a given QH liquid
does have different edge states, then we can put different edge
potentials on different segments of the edge, leading to different
edge states on different segments.  Thus two different edge states of
a given QH liquid can always be connected together.
This motivates us to ask the following question: when can we connect two
Abelian edges
$(K_1,\v q_1, \Gamma_{c1}, \Delta_1)$ and
$(K_2,\v q_2, \Gamma_{c2}, \Delta_2)$
at a point $x=0$?

Again we can fold the edge
$(K_2,\v q_2, \Gamma_{c2}, \Delta_2)$ in $(-\infty, 0)$ on top of
the edge $(K_1,\v q_1, \Gamma_{c1}, \Delta_1)$ in $(0,\infty)$.
The resulting edge is described by
$(\t K,\t {\v q}, \t \Gamma_c, \t \Delta)$:
\begin{eqnarray}
 \t K &=& \left(\matrix{ K_1&0\cr0&-K_2}\right) \nonumber\\
 \t \Delta &=& \left(\matrix{\Delta_1&0\cr0&\Delta_2}\right) \nonumber\\
 \t \Gamma_c &=& \Gamma_{c1} \oplus \Gamma_{c2} \nonumber\\
 \t {\v q} &=& \left(\matrix{ {\v q_1}\cr{\v q_2} } \right). \nonumber
\end{eqnarray}
The two edges $(K_1,\v q_1, \Gamma_{c1}, \Delta_1)$ and
$(K_2,\v q_2, \Gamma_{c2}, \Delta_2)$ can be joined together only if
the edge $(\t K,\t {\v q}, \t \Gamma_c, \t \Delta)$ can be terminated.
The different ways to join the
two edges correspond to the different ways to terminate
the edge $(\t K,\t {\v q}, \t \Gamma_c, \t \Delta)$.

\section{Some simple examples}

To gain a more intuitive understanding of the results summarized above,
we would like to discuss a few simple examples.

\subsection{$\nu=1/m$ edge state coupled to superconductor}

First let us consider an edge of a $\nu = 1/m$ Laughlin state. We place a
tunneling junction to a superconducting state at $x=0$.
After the folding, we get a two-branch edge:
\begin{eqnarray}
&&\t K = \left(\matrix{ m&0\cr 0&-m}\right) ,\ \
\t {\v q} = \bpm{1\cr 1}, \nonumber\\
&&\t \Delta = \bpm{1/m&0\cr 0&1/m}.
\end{eqnarray}
The E-lattice $\t \Gamma_c$ is generated by
$\v n^T = (m,0)$ (which creates an
electron) and $\v n^T = (1,1)$ (which transfers a quasiparticle of charge
$1/m$ from one edge to the other).
That is, $\t \Gamma_{c} = \Latt(C)$, $C=\bpm{m&1\cr0&1}$.
To understand the fixed points of the tunneling junction, we first study
the termination of the two-branch edge.

One termination (called fixed point A) is described by
$B = \bpm{1\cr 1}$. Such a termination conserves electric charge,
since $B$ satisfies (\ref{BCh}). To obtain the boundary quasiparticle
operators,
we need to find the boundary Q-lattice $\Gamma_{q_B}$, which is given by
(\ref{GaqB}).
We find
\begin{equation}
\Gamma_{q_B} = \Latt (\bpm{m/2 \cr -m/2}).
\end{equation}
An element $\v l$ of the lattice $\Gamma_{q_B}$ has the form
\begin{equation}
 \v l = l \bpm {m/2\cr-m/2}
\end{equation}
where $l \in Z$.
Thus the boundary quasiparticle operator is labeled by a single integer $l$.
We denote such an operator by $V^b_{l}$.
The scaling dimension of $V^b_{l}$ is
\begin{equation}
h^b(l) =\v l^T \t K^{-1} B (B^T \Delta B)^{-1} B^T \t K^{-1}\v l
= ml^2/2,
\end{equation}
and the charge is
$Q= \t {\v q}^T \t K^{-1}\v l = l$. It is clear
that the boundary vertex operator $V^b_{1}$ just creates an electron.
It is also clear that $V^b_{1}$ cannot appear in the boundary hamiltonian,
since only pairs of electrons can be added or subtracted from the
superconductor.

To determine the boundary operators that can appear in the
Hamiltonian, we need to consider charge conservation. In this
problem the charge ($Q$ mod 2) is conserved. Thus only $V^b_l$
with even $l$ can appear in the boundary Hamiltonian.  The leading
operator that can appear in the boundary Hamiltonian is $V^b_{2}$,
which has scaling dimension $h = 2m$ and a charge $Q = 2$. $V^b_{2}$
adds two electrons to the boundary and describes the tunneling to the
superconductor.  The scaling dimension of $V^b_2$ determines the
stability of the fixed point A.
Since the scaling dimension $2m > 1$, the fixed point is stable.
Physically, the fixed point A corresponds to a junction in which the
tunneling to the superconductor vanishes at low energies.

Another termination (called fixed point B) is described by
$B= \bpm {m\cr -m}$. Such a fixed point does not conserve the charge.
The boundary Q-lattice $\Gamma_{q_B}$ is
\begin{equation}
 \Gamma_{q_B} = \Latt( \bpm{ 1/2\cr1/2}).
\end{equation}
The elements $\v l$ in $\Gamma_{q_B}$ have the form
\begin{equation}
 \v l = l \bpm{ 1/2\cr 1/2},
\end{equation}
and the boundary quasiparticle operator is labeled by an integer $l$.
The scaling dimension of $V^b_{l}$ is
\begin{equation}
h^b = l^2/2m.
\end{equation}
Note that although the charge $Q=Q_1+Q_2$ is not conserved,
the charge difference $Q_d = Q_1-Q_2$ is conserved at low energies.
The $Q_d$ charge of $V^b_{l}$ is given by
\begin{equation}
 Q_d = \v q_d^T \t K^{-1} \v l = l/m
\end{equation}
where $\v q_d^T =(1,-1)$. This means that the boundary operator
$V^b_{l}$ transfers $l/m$ charges between the two branches.

The leading operator that can appear in the boundary Hamiltonian
is $V^b_1$, since we allow any amount of charge
to be transferred between the two
branches.  Its scaling dimension is $h^b = 1/2m$.
Since $1/2m < 1$ the fixed point B is unstable.
Physically the fixed point B corresponds to a junction with strong
tunneling to the superconductor. A low energy incoming electron will be
scattered into an outgoing hole by the junction.

Let us start with a $\nu=1/m$ edge state coupled strongly to a
superconductor at a point. At high energy scales (temperature or
applied voltage), the junction is close to the fixed point B. As the
energy is lowered, the coupling strength flows to zero and the
junction flows from the unstable fixed point B to the stable fixed
point A.  Note that the scaling dimensions of leading boundary
operators at the fixed points A and B are given by $2m$ and $1/2m$,
which are inverse of each other.  Thus the fixed point A and B form a
duality pair.  The crossover between fixed points A and B can be
solved exactly and has been studied extensively in tunneling between
FQH edge states\cite{fendley}.  In section IV we show how a
simple model incorporating interactions between electrons on different
edges can be mapped onto this exact solution.

\subsection{$(\nu_1,\nu_2)=(1/m_1, 1/m_2)$ edge state coupled to
superconductor}

Second, let us consider
a two-branch edge described by
\begin{equation}
\t K = \bpm{m_1&0\cr 0&-m_2},\ 
\t {\v q} = \bpm{ 1\cr 1} .
\end{equation}
A superconductor covers and couples to the edge in the region
$(-\infty,0)$.
We would like to ask: can the coupling produce a gap in the  $(-\infty, 0)$
region, or equivalently: can the edge terminate at $x=0$.
We find that if $\frac{m_2}{m_1}$ is a square of a rational number, then
the $B$ matrices exist and the edge can be terminated. These edges
include $(m_1, m_2)=(m,m)$, $(1,9)$, $(3,27)$,
$(9,25)$, etc. Note that the $(m_1,m_2)=(1,9)$ edge is the edge of the
$\nu=8/9$ QH state. Thus the gapless edge excitations of a $\nu=8/9$
QH state can disappear when coupled to a superconductor.

\begin{figure}
\epsfxsize=2.5truein
\vbox{\centerline{\epsffile{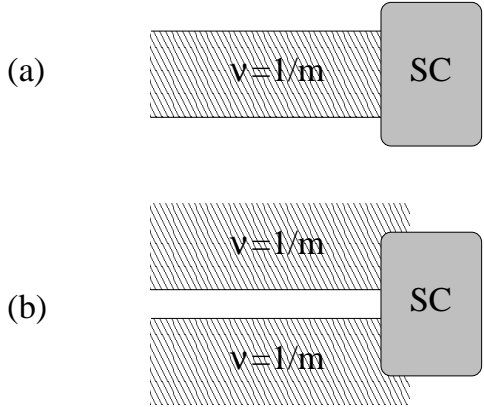}}
\caption{Two schematic setups for a superconducting contact at the
termination of two edge states.  In (a) the edge states belong to the
same droplet, in (b) to separate droplets.}
\label{figtwo}}
\end{figure}

First, we discuss the case $(m_1,m_2)=(m,m)$. This case was discussed above,
except that the lattice $\t \Gamma_c$ is different for the
present setup.  Even here there are two different possibilities.
In the first setup where the two edges belong to the same QH liquid
(Fig. \ref{figtwo}a), $\t \Gamma_c$ is given by
\begin{equation}
\t \Gamma_c = \Latt(\bpm{ 1 & m\cr 1 & 0}),
\label{GcGcB1}
\end{equation}
since we can transfer quasiparticles of charge $1/m$ between the two
branches of the edge.
This setup is identical to the case discussed in the last subsection, and
here we just repeat the results obtained before.
The fixed point A is described by $B = \bpm{1\cr 1}$.
The leading boundary operator that can appear in the
hamiltonian has scaling dimension $2m$.
The fixed point B is described by $ B = \bpm{m\cr -m }$.
The leading boundary operator that can appear in the
hamiltonian has scaling dimension $1/2m$.

In the second setup where the two edges belong to two different QH liquids
(Fig. \ref{figtwo}b), $\t \Gamma_c$ is given by
\begin{equation}
\t \Gamma_{c} = \Latt(\bpm{m & m\cr m & 0})
\label{GcGcB2}
\end{equation}
since we can only transfer electrons between the two
edge branches, not quasiparticles.
Here we can define two charges. The first one is the total electric
charge $Q$ described by the charge
vector $\t {\v q}$. Only ($Q$ mod 2) is conserved.
The second one is the difference of the electric charge on the two
branches $Q_d =Q_1-Q_2$. $Q_d$ is described by a second charge vector
$\t {\v q}_d = \bpm{1\cr -1}$. Only $(Q_d$ mod 2) is conserved here.

The fixed point A is described by $B = \bpm {m\cr m}$.
The boundary Q-lattice $\Gamma_{q_B}$ is
\begin{equation}
 \Gamma_{q_B} = \Latt( \bpm {1/2 \cr -1/2 } ).
\end{equation}
The boundary quasiparticle operator is labeled by $l$ with
$ \v l = l \bpm{1/2\cr -1/2}$.
The scaling dimension of $V^b_l$ is
\begin{equation}
h^b = l^2/2m,
\end{equation}
and total charge $Q$ is conserved at low energies.
The boundary quasiparticle operator $V^b_l$
carries definite charge $Q = \t {\v q}^T \t K^{-1}\v l =  l/m $.
The leading boundary quasiparticle operator that can appear in the boundary
hamiltonian is $V^b_{2m}$ which carries charge $Q$ mod 2 = 0 mod 2,
and has scaling dimension $2m$.

The fixed point B is described by $B = \bpm{ m\cr -m}$.
The boundary Q-lattice $\Gamma_{q_B}$ is
\begin{equation}
 \Gamma_{q_B} = \Latt( \bpm{ 1/2 \cr 1/2} ).
\end{equation}
The boundary quasiparticle operator is labeled by $l$ with
$ \v l = l \bpm{ 1/2\cr 1/2}$.
The scaling dimension of $V^b_l$ is
\begin{equation}
h^b = l^2/2m,
\end{equation}
For the fixed point B, the charge $Q_d$ is conserved at low energies.
The boundary quasiparticle operator $V^b_l$
carries definite $Q_d$ charge: $Q_d = \t {\v q_d}^T \t K^{-1}\v l =  l/m $.
Such an operator transfers $l/2m$ charges between the two branches.
The leading boundary quasiparticle operator that can appear in the boundary
hamiltonian is $V^b_{2m}$ which carries charge $Q_d$ mod 2 = 0 mod 2,
and has scaling dimension $2m$.

In the above, we see that not every boundary operator in $\t \Gamma_b$
can appear in the Hamiltonian.  One may wonder what is the meaning of
the operators which cannot appear in the Hamiltonian.  We notice that
although there are no gapless excitations in the region $(-\infty,
0)$, there are degenerate ground states in that region.  Thus there
are domain-wall-like excitations in $(-\infty, 0)$ region. The ground
state changes from one to another as we go across a domain wall. A
domain-wall-like excitations is labeled by a vector $\v l$ in the
boundary Q-lattice $\Gamma_{q_B}$. It carries the same quantum numbers
as the boundary vertex operator $V^b_{\v l}$ labeled by the same
vector $\v l$.  As we move a domain-wall like excitation to the
boundary, it becomes a boundary quasiparticle excitation described by
$V^b_{\v l}$.

\begin{figure}
\epsfxsize=2.5truein
\vbox{\centerline{\epsffile{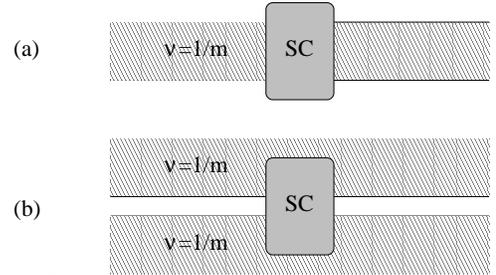}}
\caption{Two schematic setups for a superconducting contact to two
edge states.  In (a) the edge states belong to the same droplet, in (b)
to separate droplets.}
\label{figthree}}
\end{figure}

To physically measure the effect of the boundary quasiparticle excitation,
let us consider the two setups in Fig. \ref{figthree} where only a
segment of edge is covered by superconductor (The two setups in Fig.
\ref{figthree} are
related to the two
setups discussed above). In those setups, we can create a pair of boundary
excitations: the first one is
 created by $V^b_{\v l}$, $\v l \in \t \Gamma_b$ at one boundary
and the second one is created by $V^b_{-\v l}$ at the other boundary.
Such an operation transfers a domain-wall-like excitation from one boundary
to the other, and is an allowed process.
For the fixed point A in the first setup,
the lowest scaling dimension for the above
pair-creation operators is $m$.
For the fixed point B in the first setup,
the lowest scaling dimension for the
pair-creation operators is $1/m$.
For $m>1$, the fixed point A is stable while the fixed point B is unstable.
For both fixed point A and B in the second setup,
the lowest scaling dimension for the
pair-creation operators is $1/m$. For $m>1$,
the operator is relevant (since $1/m < 1$), and
the two edge states separated by the superconductor will join together at
low energies.

Next, we
would like to discuss the $(m_1,m_2)=(1,9)$ case.
We will also consider the possibility of short-ranged density-density
interactions between modes on the $m_1 = 1$ and on the $m_2 = 2$ edge.
The interaction effects being discussed are described in detail in
the next section for a different case; to avoid duplication we will
simply start here from the boost form (\ref{boost}) of the matrix
$\t \Delta$:
\begin{equation}
\t \Delta = \bpm{1&0\cr 0&1/3} \bpm{\cosh \tau & \sinh \tau \cr
\sinh \tau & \cosh \tau} \bpm{1&0\cr 0&1/3}.
\end{equation}
The boost parameter $\tau$ measures the interaction strength:
$\tau = 0$ corresponds to unmixed $\nu = 1$ and $\nu = 1/9$ states.
The lattice for charge excitations $\t \Gamma_c$ is given by
\begin{equation}
\t \Gamma_{c} = \Latt(\bpm{1&0\cr 0&9}).
\end{equation}
The first termination (called fixed point A) is described by
$B= \bpm{3 \cr 9}$.
The boundary Q-lattice is given by (\ref{GaqB}):
\begin{equation}
 \Gamma_{q_B} = \Latt( \bpm{1/6\cr -1/2 } )
\end{equation}
The boundary quasiparticle operators are labeled
by an integer $l$ with $\v l= l \bpm{1/6\cr -1/2}$.
The boundary operator $V^b_l$ has scaling
dimension
$h^b = \frac{l^2}{36} (\cosh \tau - \sinh \tau)$.
The combined charge $Q_{(1,3)} = Q_1+3 Q_2$ described by the vector
$\v q_{(1,3)}^T=(1,3)$ is conserved at low energies.
The $Q_{(1,3)}$ charge of $V^b_l$ is
$Q_{(1,3)} = \v q_{(1,3)}^T \t K^{-1}\v l = l/3$.
The leading operator that can appear in the Hamiltonian is
$V^b_6$ which has an even number of $Q_{(1,3)}$ charges and a scaling
dimension $h^b = \cosh \tau - \sinh \tau$.

The second termination (called fixed point B) is described by a lattice
 $B= \bpm{3 \cr -9}$. The boundary Q-lattice is given by
\begin{equation}
 \Gamma_{q_B} = \Latt( \bpm {1/6 \cr 1/2} ).
\end{equation}
The boundary quasiparticle is labeled by an integer $l$ with
$\v l= l  \bpm{1/6\cr 1/2}$. The boundary operator $V^b_l$ has scaling
dimension $h^b = \frac{l^2}{36} (\cosh \tau + \sinh \tau)$.
The combined charge $Q_{(1,-3)} = Q_1-3Q_2$ described by the vector
$\v q_{(1,-3)}^T=(1,-3)$ is conserved at low energies.
The $Q_{(1,-3)}$ charge of
$V^b_l$ is $Q_{(1,-3)} = \v q_{(1,-3)}^T \t K^{-1}\v l = l/3$.
The leading operator that can appear in the Hamiltonian is
$V^b_6$ which has an even number of $Q_{(1,-3)}$ charges and a scaling
dimension $h^b = \cosh \tau + \sinh \tau$.  Note that since
$(\cosh \tau - \sinh \tau) (\cosh \tau + \sinh \tau) = 1$, fixed
points A and B form a duality pair.

\subsection{Junction between $\nu_1 = 1/m_1$, $\nu_2 = 1/m_2$ edge
states}

This case has been treated previously in the absence of intermode
interactions by several authors~\cite{kane1,fendley,chk,chamon}.
Here we will show how the known results for the two fixed points
are recovered in our framework, and in the next section consider
interaction effects.  A new result even in the absence of interactions
is the information found below about the lattice of charged boundary
operators.  For definiteness we consider the case of tunneling between
$\nu = 1$ and $\nu = 1/3$ states which has attracted the most attention.
At weak coupling between the two edges, the most relevant neutral operator
tunnels an electron between the two edges, with scaling dimension 2; at
strong coupling the most relevant neutral operator has scaling dimension
$\frac{1}{2}$.  This neutral operator can be interpreted as 
tunneling between different minima of the boundary cosine interaction
in the sine-Gordon model, or as quasiparticle tunneling in an effective
model of two $\nu=\frac{1}{2}$ edges.

With no intermode interactions, we have
\begin{eqnarray}
&&\t K = \bpm{m_1&0&0&0\cr 0&-m_1&0&0\cr
0&0&m_2&0\cr 0&0&0&-m_2},\ 
\t {\v q} = \bpm{ 1\cr 1 \cr 1 \cr 1}\nonumber \\
&&\t \Delta = \bpm{1/m_1&0&0&0\cr 0&1/m_1&0&0\cr
0&0&1/m_2&0\cr 0&0&0&1/m_2}.
\end{eqnarray}
Fixed point A (weak coupling) is described by
\begin{equation}
B = \bpm{1 &0\cr 1&0\cr 0&3\cr 0&3}.
\end{equation}
The meaning of this $B$ matrix is that the incoming and outgoing
$\nu = 1$ edges are joined continuously to each other, and similarly
for the incoming and outgoing $\nu = 1/3$ edges.
The boundary Q-lattice is
\begin{equation}
\Gamma_{q_B} = \Latt(\bpm{\frac{1}{2}&0\cr-\frac{1}{2}&0\cr0&\frac{1}{2}\cr
0&-\frac{1}{2}}).
\end{equation}
A general boundary operator $O_{\v l}$
with $\v l = l_1 (\frac{1}{2},-\frac{1}{2},0,0)$
$+ l_2 (0,0,\frac{1}{2},-\frac{1}{2})$, $l_1$ and $l_2$ integers,
has electric charge $Q_{\v l} = l_1 + l_3 / 2$ and scaling dimension
${l_1}^2 / 2 + {l_2}^2 / 6.$  The most relevant neutral
boundary operator has $(l_1,l_2) = (1,-3)$ and scaling dimension 2,
as expected for electron tunneling at weak coupling.

Fixed point B (strong coupling) is described by
\begin{equation}
B = \bpm{2 &1\cr 1&1\cr 0&1\cr 3&1}.
\end{equation}
The meaning of the first column of $B$ is that two incoming electrons
become one outgoing electron and three outgoing $q=\frac{1}{3}$.
quasiparticles.  In the boundary sine-Gordon language, this corresponds
to pinned (Dirichlet) boundary condition on the neutral mode.
The second column is simply the charge vector, indicating
that overall charge is conserved across the junction.
The boundary Q-lattice is
\begin{equation}
\Gamma_{q_B} = \Latt(\bpm{{3 \over 10}&-{1 \over 10}\cr
{3 \over 10}&-{4 \over 5}\cr
-\frac{7}{10}&\frac{3}{2}\cr
-\frac{7}{10}&\frac{3}{5}}).
\end{equation}
A general boundary operator $O_{\v l}$
with $\v l = l_1 (\frac{3}{10},\frac{3}{10},-\frac{7}{10},-\frac{7}{10}) +
l_2 (-\frac{1}{10},-\frac{4}{5},\frac{3}{2},\frac{3}{5})$ has charge
$Q = l_2$ and scaling dimension ${l_1}^2 / 2 - 3 l_1 l_2 / 2 +
3 {l_2}^2/2.$  The most relevant neutral boundary operator has
$(l_1,l_2) = (1,0)$ and scaling dimension $\frac{1}{2}$, as expected.


\section{Effects of short-range interactions}

In this section we consider the effect of short-range interactions on
tunneling through a point contact between two Laughlin states.  In the
absence of interactions, the nonlinear $I-V$ curve was found
exactly by Fendley, Ludwig, and Saleur~\cite{fendley} via a mapping
onto the integrable boundary sine-Gordon (BSG) model.  First we show in the
BSG formalism that a simple solvable model
incorporating interactions gives a continuous renormalization of the
effective fractional charge appearing in the $I-V$ characteristic.
We use the BSG formalism since we will eventually be interested not
only in the fixed points, which can equally well be described in
the $B$-matrix formalism of section III,
but also in the crossover.
The $I-V$ curve measured in tunneling experiments in real systems,
where the screened Coulomb interaction is present, will thus be
sensitive in some geometries (discussed below) to nonuniversal
electron-electron interactions.  Our model is different from that
of Pryadko \etal~\cite{pryadko}, which uses a long-range Coulomb interaction
regularized by an opening angle at the junction.

The effective action describing tunneling between edges of two Laughlin
states with filling fractions $\nu_1 = 1/m_1$, $\nu_2 = 1/m_2$ is
\begin{eqnarray}
S &=& S_{\rm free} + S_{\rm tun},\cr S_{\rm free} &=& {1 \over 4 \pi}
\int dx\,dt\,\sum_{ij} [ K_{ij} \partial_t \phi_i \partial_x \phi_j -
V_{ij} \partial_x \phi_i \partial_x \phi_j],\cr S_{\rm tun} &=& \Gamma
\delta(x) (e^{i m_1 \phi_1 - i m_2 \phi_2}).
\label{chill}
\end{eqnarray}
Here the matrix $K$, which describes the statistics of the vertex operators
$e^{i n_i \phi_i}$
created from the bosonic fields, is
\begin{equation}
K = (\matrix{m_1 & 0\cr 0 & -m_2})
\end{equation}
where we have taken the two edges to propagate in opposite directions.
If $V$ is diagonal the physics is independent of whether the modes are
copropagating or counterpropagating, but we are interested in the case
of general $V$ in which case there are differences, as seen below.  If
the positive definite matrix $V$ is diagonal, its two entries $V_{11}$
and $V_{22}$ are the velocities of the two modes.  The off-diagonal
elements of $V$ correspond to a density-density interaction across the
two edges, since the electron density is proportional to $\partial_x
\phi_i$ for each mode.

The above action maps onto a boundary sine-Gordon model, with boson radius
determined by the filling fractions of the original states and by the
matrix $V$.  The boundary sine-Gordon model contains one nonchiral boson
(i.e., with both left and right components) on the half-line.  The mapping
consists of rotating the fields $\phi_1, \phi_2$ so that one new combination
${\tilde \phi}_1$ is proportional to the exponent $m_1 \phi_1 - m_2 \phi_2$
in $S_{\rm tun}$, while ${\tilde \phi}_2$ does not appear in $S_{\rm tun}$
and hence is free.  Then folding the field ${\tilde \phi}_1$ onto the half-line
and rescaling gives the action
\begin{equation}
S_{\rm BSG} = \int dt \int_{-\infty}^0 dx [{(\partial_x \Phi)^2 \over 2}
+ {(\partial_t \Phi)^2 \over 2} +
\cos(\beta \Phi / 2)].
\label{BSG}
\end{equation}
The constant $\beta$, given for diagonal $V$ by 
$\beta = \sqrt{4 \pi / (\frac{1}{\nu_1} + \frac{1}{\nu_2})}$,
gives the tunneling term in (\ref{BSG}) the same scaling
dimension $\Delta = ({m_1}^2 + {m_2}^2)/2$ as in (\ref{chill}).
The velocities of the edge modes, defined as the velocities in a
basis where $V$ is diagonal, should strictly speaking be equal for this
rotation of fields to be valid, but since the tunneling takes place
at a point and there is no coherence along the edge, a difference in
velocities should not have much effect.

In order to calculate the conductance across the tunneling junction,
the effective $\beta$ which appears in $S_{\rm BSG}$ needs to be
determined, as well as the contribution $q_{\rm eff}$ to the current
from each tunneling event.  Previously only certain discrete values of
$\beta$, corresponding to tunneling between Laughlin states, were thought to
be physically relevant for edge tunneling.  This is because a general
$\beta$ describes tunneling between two chiral Luttinger liquids with
continuous Luttinger parameter, but only specific values of the
Luttinger parameter correspond to quantum Hall states $\nu = 1/m$.
The main result of this section is that tunneling between Laughlin
states {\it with non-diagonal $V$} is described by the boundary
sine-Gordon model with continuously varying $\beta$
and $q_{\rm eff}$.

\begin{figure}
\epsfxsize=2.5truein
\vbox{\centerline{\epsffile{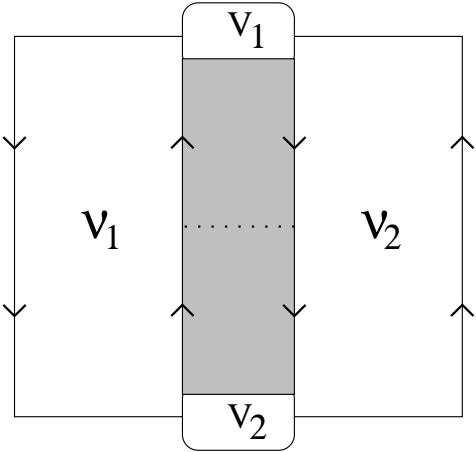}}
\caption{Possible experimental geometry for point
tunneling between quantum Hall states $\nu_1$ and $\nu_2$.
The density-density interaction between edges
is nonzero in the shaded region, and zero elsewhere.  The two contacts
at voltages $V_1$ and $V_2$ are assumed to populate edge modes propagating
away from the contact up to energy $e V$.}
\label{figfour}}
\end{figure}

The model which we solve exactly has a region of constant interaction
strength (between contacts $V_1$ and $V_2$ in Fig.~\ref{figfour}) and
zero interaction elsewhere.  It is essential that the two modes in the
interaction region be oppositely directed, so that the scaling dimension
of the tunneling operator is affected by $V$.  The first step is to write the
positive definite matrices $V$ and $\Delta$ in terms of a ``boost''
parameter $\tau$.  The advantage of doing so is that $\Delta$ is only
a function of $\tau$ and not of the eigenmode velocities $v_i$ which
affect $V$; the boost decomposition~\cite{moorewen} isolates the
dependence of $\Delta$ on as few parameters as possible.
\begin{eqnarray}
V &=& K^{1/2} B
\left(\matrix{v_1 & 0 \cr 0 & v_2}\right)
B K^{1/2},\cr
\Delta &=& K^{1/2} B
\left(\matrix{1 & 0 \cr 0 & 1}\right)
B K^{1/2},\cr
K^{1/2} &=& \left(\matrix{\sqrt{m_1} & 0 \cr
0 & \sqrt{m_2}}\right),\cr
B &=& \left(\matrix{\cosh \tau & \sinh \tau \cr
\sinh \tau & \cosh \tau} \right).
\end{eqnarray}
Now the scaling dimension and transferred charge of the tunneling
operator can be simply expressed in terms of $\Delta(\tau)$.  In the
following we will specialize to the case $m_1 = 1$, $m_2 = 3$, i.e., a
$\nu_1 = 1$ edge and $\nu_2 = \frac{1}{3}$ edge propagating in
opposite directions.  The results generalize simply to other values of
$m_1$ and $m_2$, although the boundary sine-Gordon crossover result
used below requires that at most one relevant operator be
present, restricting $m_1$ and $m_2$ somewhat~\cite{fendley}.

The intuitive meaning of the effective charge transfer can be
understood by considering the ``charge-unmixed'' point
$\tau = 0$ where one of the two eigenmodes (i.e., modes
which diagonalize $K$ and $V$) is neutral, and one charged.  Then
in this basis the tunneling operator, which is neutral, is of
the form $\exp(i C \phi_{n}) + {\rm h.c.}$ for some constant $C$.
The tunneling operator for this value of $\tau$ does not transfer
any charge from one eigenmode to the other, since only one
eigenmode carries charge.  Hence the conductance measured across
the junction is {\it independent} of the rate of tunneling events
determined by the coefficient of the tunneling operator.
The value $\tau = \log(1+\sqrt{3}/\sqrt{2})$ corresponds to
decoupled $\nu = 1$ and $\nu = 1/3$ edges in the interaction region,
which has been previously studied in a number of
works.~\cite{kane,chk,chamon}

The scaling dimension of the tunneling operator $\exp(i m_i \phi_i)$
is $\frac{1}{2} {\bf m} \Delta(\tau) {\bf m}$.  The effective
tunneling charge is determined by how far the neutral tunneling
operator ${\bf m}$ is from being an eigenmode of the system:
$q_{\rm eff} = \frac{e}{2} {\bf q} \Delta {\bf m}$,
which is zero if ${\bf m}$ is an eigenmode and
grows as $\Delta$ moves away from the charge-unmixed point.
The current measured across the two contacts of
Fig.~\ref{figfour} if there is no electron tunneling across the
junction (if $V_1 - V_2$ is small) is
$I = (V_1 - V_2) (\frac{2}{3} + \frac{1}{2}(\sigma_1 + \sigma_2))$,
where $\sigma_1$ and $\sigma_2$ are the conductances along
the edges in the interaction region of Fig.~\ref{figfour}.
We use the identity $\sigma_1 + \sigma_2 = {\bf t} \Delta {\bf t}$
in what follows.
 
The total change in conductance from small $V$ to large $V$ is fixed
by the scaling dimension of the electron tunneling operator and the
effective tunneling charge.  From previous work~\cite{schmid}, it is
known that if the effective tunneling charge is 1, the conductance
change is ${e^2 \over h \Delta}$, where $\Delta$ is the scaling
dimension of the electron tunneling operator.  For instance, in
tunneling between two $\nu = 1/3$ states, the conductance change
between no tunneling (two Hall droplets)
and no backscattering (one Hall bar) is ${e^2 \over 3 h} =
{e^2 \over h \Delta}.$  In the case of interactions, $e$ must be replaced
by the effective charge transfer per tunneling event $q_{\rm eff}$.
Hence for the system of Fig.~\ref{figfour},
\begin{eqnarray}
{h \sigma_{\rm max} \over e^2} &=&
\frac{2}{3} + {{\bf t} \Delta {\bf t} \over 2} =
{2 + \cosh(2 \tau) \over 3},\cr
{h \sigma_{\rm min} \over e^2} &=&
\frac{2}{3} + {{\bf t} \Delta {\bf t} \over 2}
- {({\bf t} \Delta {\bf m})^2 \over 2 {\bf m} \Delta {\bf m}} \cr
&=& {2 + {\rm sech}(2 \tau) \over 3}.
\label{minmax}
\end{eqnarray}

In fact the whole conductance curve between these two values can be
calculated from the mapping to $S_{\rm BSG}$.  Before doing so, there
is a simple check on our results which gives some insight into why the
above values are natural.  Assuming conservation of energy (i.e., no
dissipation at the junction~\cite{msc}) gives two possible values of
the current: $I = (V_1 - V_2) (\frac{2}{3} + \frac{1}{2}(\sigma_1 +
\sigma_2))$, corresponding to no tunneling current, and $I = (V_1 -
V_2) ({2 \over 3} + {2 \over 9 (\sigma_1 + \sigma_2)}).$ The two
corresponding values of the conductance are exactly those in
(\ref{minmax}).  Thus our calculation reproduces the asymptotic values
of the conductance consistent with zero-dissipation fixed points.

\begin{figure}
\epsfxsize=3.0truein
\vbox{\centerline{\epsffile{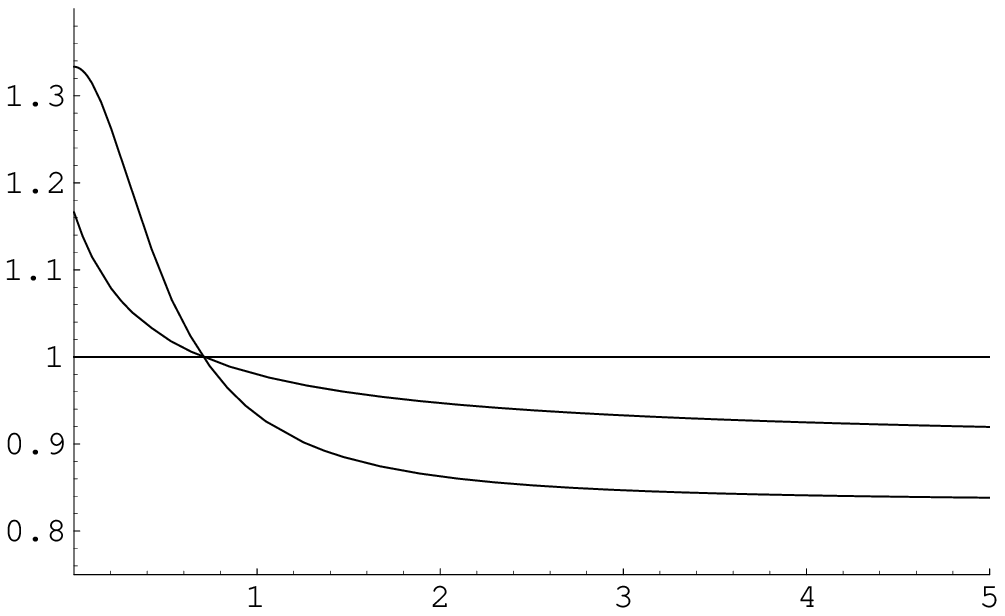}}
{\vskip -0.9 in
\noindent ${dI \over dV}$ \vskip 0.9 in}
{\vskip -1.14 in
\hbox{\hskip 2 in \parskip=2 pt
\vbox{$\tau = 0$ \par
$\tau = \log(1 + \sqrt{3} / \sqrt{2})$ \par
$\tau = \log(1 + \sqrt{5} / 2)$}}
\vskip 0.5 in
\centerline{$C(\tau) V$}\vskip 0.1 in}}
\caption{Differential conductance $dI/dV$ in units of $e^2 / h$
versus scaled voltage $C(\tau) V$
for three different values of interaction strength (boost parameter)
$\tau$.  The horizontal axis is expected to scale by a different
cutoff-dependent constant $C(\tau)$ for each value of $\tau$.}
\label{figfive}
\end{figure}

The current-voltage characteristic can be calculated
(Fig.~\ref{figfive} up to one overall constant in the energy scale,
which corresponds to the initial strength of tunneling.~\cite{fendley}
The result is, with $V = V_1 - V_2$ and $\sigma_{\rm max}$ as in
(\ref{minmax}),
\begin{eqnarray}
I &=& \sigma_{\rm max} V - I_{\rm tun},\cr
I_{\rm tun} &=& \cases{I^{(1)} & if
$V < T_B \Delta^{-1/(2-2/\Delta)} \sqrt{\Delta-1}$  \cr
I^{(2)} & if $V \geq T_B \Delta^{-1/(2-2/\Delta)} \sqrt{\Delta-1}$},\cr
I^{(1)} &=& {q_{\rm eff}^2 V \over h}
\sum_{n=1}^\infty f_n(\Delta),\cr
I^{(2)} &=& {q_{\rm eff}^2 V \over h \Delta}
\left(1 - \sum_{n=1}^\infty {f_n(1/\Delta) \over \Delta}\right),\cr
f_n(g) &=& {(-1)^{n+1} \sqrt{\pi} \gamma(n g) \over
2 \Gamma(n) \Gamma(3/2 + n(g - 1))}
\left({e V \over {T_B}^\prime}\right)^{2n (g - 1)}.
\end{eqnarray}
Here $T_B$ is some cutoff-dependent constant which may vary with the
boost parameter $\tau$.  This calculated current should be relevant
as long as the interaction strength is nearly constant in the region
around the tunneling junction.  The details of the interaction far away
from the tunneling junction should not matter as long as current is
conserved in the incoming and outgoing edge branches.

\section{Correlation functions in bounded Luttinger liquids}

\subsection{Review of Abelian edge states}

First we review the $\chi$LL description of the
edge of an Abelian QH state.~\cite{wenrev0}
The collective modes that propagate along the edge
can be described by several density operators $\rho_a(x)$, which
satisfy the algebra
\begin{equation}
 [\rho_{a}(x), \rho_{b}(y) ] = \frac{i}{2\pi}
 (K^{-1})_{ab} \delta^\prime (x-y)
\end{equation}
where $K$ is a symmetric integral matrix that characterizes an Abelian edge.
The total electric density is a sum of $\rho_a$ weighted by a charge vector
$q^a$:
\begin{equation}
 \rho_e = q^a \rho_a.
\end{equation}
If there is no binding between electrons, we can choose the symmetric basis
in which $q^a = 1$. In the following we will work in this symmetric basis.

The edge dynamics is described by the Hamiltonian
\begin{equation}
 H = \int dx\, \pi V_{ab} \rho_{a}(x) \rho_{b}(x)
\end{equation}
at low energies, where $V$ is positive definite symmetric matrix.
The corresponding Lagrangian density is given by
\begin{equation}
 \cL =-\frac{1}{4\pi}\left( K^{ab} \phi_a \partial_t \partial_x\phi_b
-V^{ab} \phi_a \partial_x^2\phi_b \right)
\label{Ledge}
\end{equation}
where $\partial_x \phi_a/2\pi = \rho_a$.

For a circular edge (of length $L$), if we assume $\phi_a$ to satisfy
the periodic boundary condition
$\phi_a(0) = \phi_a(L)$, then the $\v \phi$ only
describe the neutral fluctuations. To go beyond the neutral fluctuations,
let us first consider the charge fluctuations created by
electron operators $\psi_e$. There are $\dim(K)$ different electron
operators
\begin{equation}
 \psi_e^b(x) = e^{i n_e^{ab} \phi_a(x)}
\end{equation}
where $\v n_e^b$ is the $b^{th}$ column of $K$: $n_e^{ab} = K^{ab}$.
An multi-electron operator has a form $e^{i n_e^{a} \phi_a(x)}$ with
$\v n_e$ belonging to a lattice $\Gamma_e$, which will be called the electron
lattice.  The lattice $\Gamma_e = \Latt(K)$,
since $\Gamma_e$ is generated by the column vectors of $K$.

The charge fluctuations created by the
electron operators can also be described by
the $\v \phi$ fields in the Lagrangian. But now the $\v \phi$ fields
no longer satisfy the periodic boundary condition. Instead they satisfy
\begin{equation}
 \phi_a(L) = \phi_a(0) + 2\pi N_a
\end{equation}
where $N_a$ are integers. In fact, the above $\v \phi$ describe excitations
created by $\prod_b (\psi_e^b)^{N_b}$ from the neutral ground state.

In general, the (multi-)electron operators do not represent all possible
``charge'' excitations. The most general charge excitations are created by
\begin{equation}
\psi_c(x) = e^{i n_c^{a} \phi_a(x)}
\end{equation}
where $\v n_c$ are integral vectors belonging
to a lattice $\Gamma_c$ (which will
be called the E-lattice). The general
charge excitations contain all excitations created by the electron
operators,
and thus $\Gamma_e \subseteq \Gamma_c$. The general
charge excitations can also include excitations that transfer
fractional charges between different edge branches, but have
integer overall charge.

The E-lattice $\Gamma_c$ must satisfy certain conditions.
Since $\psi_c(x)$ all carry integral charges (which can be zero), the
vectors
in $\Gamma_c$ satisfy
\begin{equation}
 \v q^T K^{-1} \v n_c = \hbox{integer}
\end{equation}
The operators $\psi_c(x)$ are also mutually local; that is
\begin{equation}
 (\v n_c^\prime)^T K^{-1} \v n_c = \hbox{integer}
\end{equation}
for any $\v n_c$ and $\v n_c^\prime$ in $\Gamma_c$.
This condition implies that $\Gamma_c \subseteq \Latt(K^{-1})$.
Just like the electron operators, the excitations created by
$\psi_c=e^{i\v n_c\cdot \v \phi}$ are
described by $\v \phi$ that satisfy the boundary condition
\begin{equation}
 \v \phi(L) = \v \phi(0) + 2\pi K^{-1} \v n_c
\end{equation}
We can also use the above to define the periodicity of the $\v \phi$
field. That is $\v \phi$ and $\v \phi^\prime$ are equivalent if
\begin{equation}
 \v \phi = \v \phi^\prime  + 2\pi K^{-1} \v n_c
\end{equation}
for some $\v n_c \in \Gamma_c$.

\subsection{Correlations with boundary}

Now we add a tunnel junction or termination to the system.  In
what follows it is assumed that in the junction case
the edge has been folded onto
the half-line as in section II, so that one set
$(\t K,\t{\v q}, \t \Gamma_c,\t\Delta)$ describes both incoming
and outgoing edges, with a total of $2 k$ bosonic fields.
It will be shown below that the number of incoming fields must
equal the number of outgoing fields.
In the presence of tunneling between different edge
branches at $x=0$, the Lagrangian also contains terms of type
\begin{equation}
 \cL_t = \sum_I t_I \delta(x) e^{\v n^I\cdot \v \phi} + h.c.
\end{equation}
where $t_I$ is the (real) tunneling strength, and the integral
vectors $\v n^I$ are points in the $\Gamma_c$ lattice
which also satisfy
\begin{equation}
( \v n^I)^T {\t K}^{-1} \v n^I = \hbox{even}.
\end{equation}
(Thus $e^{i\v n^I\cdot \v \phi}$ are bosonic operators.)  If electric
charge is conserved at the boundary, $\v n^I$ should also satisfy
\begin{equation}
 \v q^T K^{-1} \v n^I =0
\end{equation}
so that $e^{i\v n^I\cdot \v \phi}$ represents a neutral operator.
For the time being, we will not impose the above charge-conservation
condition.

The Lagrangian density on the half-line is given by
\begin{equation}
 \cL =-\frac{1}{4\pi}\left( (\t K)^{ab}
\t \phi_a \partial_t \partial_x\t \phi_b
-(\t V)^{ab} \t \phi_a \partial_x^2\t \phi_b \right)
\label{LtK}
\end{equation}
We also need to specify what are the allowed
charge excitations, in addition to the Lagrangian (\ref{LtK}).
This can be achieved by specifying the periodicity conditions
of the $\t{\v \phi}$ fields:
\begin{equation}
 \t{\v \phi} \sim  \t{\v \phi}^\prime = \t{\v \phi}  + 2\pi \t K^{-1} \v n_c
\end{equation}
where $\v n_c$ are vectors in the lattice $\t \Gamma_c$.

To completely define our system on the half-line, we need to include
boundary conditions at $x=0$.
First let us consider the boundary condition in Hamiltonian language.
We will start with the following type of boundary conditions, specified
by a collection of vectors $B^a_I$ with $I=1,...,k$:
\begin{equation}
 B^a_I \t \phi_a(0) = 0,\ \ \text{for all } I
\label{Bphi}
\end{equation}
Since we are looking for critical points, we would like to study boundary
conditions that are invariant under scaling, suggesting the form (\ref{Bphi}).
However, the above form is inconsistent with the periodicity conditions on
$\t {\v \phi}$.  It turns out that (\ref{Bphi}), although sufficient to
determine the correlation functions of vertex operators,
must be improved (later in this section) in order to determine the
lattice of allowed boundary quasiparticle operators.

Since $\t \phi_a(x)$'s do not commute with each other, the above boundary
conditions are self-consistent only if
\begin{equation}
[B^a_I \t \phi_a(x), B^b_J \t \phi_b(y) ] = 0,\ \ \text{for all } I,J.
\end{equation}
This implies that the vectors that characterize the boundary condition must
satisfy
\begin{equation}
 B^T \t K^{-1} B = 0,\ \ \ \det(B^TB) \neq 0
\label{BKB}
\end{equation}
where $B$ is the $2k$ by $k$ matrix formed by $B^a_J$.
Note that two different $B$ matrices, $B_1$ and $B_2$,
specify the same boundary condition in the sense of (\ref{Bphi})
if they are related by
\begin{equation}
 B_1 = B_2U,\ \  U \in L(k)
\label{BsimB}
\end{equation}
where $U$ is an $k\times k$ invertible real matrix $U$.
In this case we say $B_1$ and $B_2$ are ``weakly'' equivalent
\begin{equation}
 B_1 \sim B_2.
\end{equation}

In the Lagrangian language, the above boundary condition corresponds
to $ B^T \t {\v \phi}(0) = 0$ on the fields $\t {\v \phi}$.  Only for
certain choices of $B$ can the operator $\cK = (\t K)^{ab} \partial_t
\partial_x -(\t V)^{ab} \partial_x^2$ be hermitian.  We will show that
the condition on $B$ that makes $\cK$ hermitian is nothing but
(\ref{BKB}). To see this, we first note that
\begin{equation}
 \int dxdt \t{\v \phi}_2^T \cK \t{\v \phi}_1
= \int dxdt \t{\v \phi}_1^T \cK \t{\v \phi}_2
+\int dt (\partial_t \t{\v \phi}_2^T K \t{\v \phi}_1)_{x=0}
\end{equation}
Thus $\cK$ is hermitian if $B$ is such that
\begin{equation}
 \t \phi_a \t K^{ab} \t \phi_b^\prime =0
\label{phiKphi}
\end{equation}
for all $\t{\v \phi}$, $\t{\v \phi}^\prime$ that satisfy
$B^T\t{\v \phi}=B^T\t{\v \phi}^\prime=0$.
Let us choose a basis that the vectors in the null space of $B$ have
the form $\t{\v \phi}^T_{null}=(0,...,0,a,...,b)$, i.e., the first $k$
elements are zero.. In this basis, by (\ref{phiKphi}) $\t K$ has the
form
\begin{equation}
 \t K = \bpm {K_{0} & K_{1} \cr K_{1}^T & 0},
\label{KK0K1}
\end{equation}
and $B$ has the form
\begin{equation}
 B = \bpm {B_{1} \cr 0 }
\end{equation}
where $K_1$ and $B_1$ are invertible.
Since $\t K^{-1}$ has the form
\begin{equation}
 \t K^{-1} = \bpm{0 & K_{1}^{-1} \cr
(K_1^T)^{-1} & -(K_1^T)^{-1}K_0K_1^{-1}},
\end{equation}
thus the condition (\ref{phiKphi}) implies the condition (\ref{BKB}).
We see that the hermiticity of the operator $(\t K)^{ab}
\partial_t \partial_x -(\t V)^{ab} \partial_x^2$ also requires $B$ to
satisfy (\ref{BKB}).

Let us summarize our results so far.  The critical boundary
conditions of an Abelian edge described by $\t K$ are characterized by a
matrix $B$ that satisfies (\ref{BKB}).
If such a $B$ does not exist,
then the edge state described by $\t K$ cannot be
terminated at a point.
If such a $B$ does exist, then we can consistently impose a
boundary condition $B^T \t{\v \phi} = 0$
and terminate the edge state at $x=0$ (provided that $B$ satisfies some
other conditions that will be discussed later).
If more then one inequivalent $B$ exists,
then the edge can be terminated in more
than one way (the tunneling junction has more than one fixed point).

Now the question is when $B$ exists.  First, from (\ref{KK0K1})
we see that the signature of $\t K$ must be zero in order for $B$ to exist.
Since $\t K$ is invertible, we can write $\t K$ as
\begin{equation}
\t K = \t K_{1/2}^T \bpm{I_{k\times k} & 0  \cr
          0   &-I_{k\times k}}  \t K_{1/2}.
\end{equation}
We see that $B$ always exists for the above $\t K$.
A generic $B$ that satisfies (\ref{BKB}) can be written as
\begin{equation}
 B= \bpm{1\cr T_0^T} \t K_{1/2}, \ \ \ T_0 T_0^T=I.
\end{equation}
Thus $B$ exists if and only if $\t K$ has a
vanishing signature (there are as many incoming modes as outgoing modes).
In this case the different boundary conditions are
labeled by an element in the $O(k)$ group.

Next we would like to calculate the correlation functions of the fields
$\t{\v \phi}$ in the presence of the boundary.
First we choose a basis in which both
$\t K$ and $\t V$ are diagonal (this can always be done \cite{wenrev0}):
\begin{equation}
 \t K = \bpm{I&0\cr0&-I} \equiv \Sigma_3,\ \ \
 \t V = \bpm{v_R&0\cr0&-v_L} \equiv \t V_{diag}
\end{equation}
The block $v_R$ has
positive diagonal elements which are the velocities of right
movers, while $v_L$ has negative diagonal elements which are the
velocities of left movers. Thus the velocity for each branch is given by
$v_a =(\Sigma_3 \t V_{diag})_{aa}.$
In this basis a generic $B$ is equivalent to a simple form:
\begin{equation}
 B \sim \bpm{I\cr T_0^T}, \ \ \ T_0 T_0^T=I
\label{simB}
\end{equation}
Let $G_{0,ab}(x,t)$ be the correlation of
$\t \phi_a(x,t)$ and $\t \phi_b(0)$ in the
absence of the boundary: $G_{0,ab}(x,t) = \vev{\t \phi_a(x,t)\t
\phi_b(0)}_0$.  The correlation satisfies a linear equation
\begin{eqnarray}
&& \frac{1}{2\pi} \left( - \t K^{ac} \partial_{x_1} \partial_t
+ \t V^{ac} \partial_x^2 \right) G_{0,cb}( x ,t) = \delta_{ab} \delta(x)\delta(t)
\end{eqnarray}
Since $\t K$ and $\t V$ are diagonal in the present basis, the above can be
rewritten as
\begin{eqnarray}
&& \frac{1}{2\pi} \left( - \t K^{aa} \partial_{x} \partial_t
+ \t V^{aa} \partial_x^2 \right) G_{0,aa}( x ,t) = \delta(x)\delta(t) \\
&& a = 1,...,\dim{\t K} . \nonumber
\end{eqnarray}
We notice that the pair $(a,x)$ is simply a label of the $\t {\v \phi}$
field.
We can choose another label $(a, \t x)$ to eliminate the velocities and
further simplify the above
equation. The two sets of labels are related by
\begin{equation}
 (a, x) = (a, \t V^{aa} \t x)
\end{equation}
Now the equations for $G_{0,aa}$ become
\begin{eqnarray}
&& \frac{1}{2\pi} \left( - \t \Sigma_3^{aa} \partial_{\t x} \partial_t
+ \partial_\t x^2 \right) G_{0,aa}( \t x ,t) = \delta(\t x)\delta(t) \\
&& a = 1,...,\dim{\t K} \nonumber
\end{eqnarray}
The correlation function in the presence of the
boundary, $ G_{ab}(x_1, x_2,t) = \vev{ \t \phi_a(x_1, t) \t \phi_b(x_2, 0)}$,
satisfies according to (\ref{Bphi})
\begin{equation}
  B^T G(0, x_2,t) = G(x_1, 0,t) B = 0,
\label{BGGB}
\end{equation}
and
\begin{eqnarray}
&& \frac{1}{2\pi} \left( - \t K^{ac} \partial_{x_1} \partial_t
+ \t V^{ac} \partial_{x_1} ^2 \right) G_{cb}( x_1,x_2 ,t) \nonumber\\
&=& \delta(x_1-x_2)\delta(t)
\label{EqG1}
\end{eqnarray}
for $x_1, x_2 > 0$.
In the diagonal basis and in terms of the new label $(a, \t x)$,
the above equation becomes
\begin{eqnarray}
&& \frac{1}{2\pi} \left( - \t \Sigma_3 \partial_{\t x_1} \partial_t
+  \partial_{\t x_1} ^2 \right) G( \t x_1,\t x_2 ,t) = \delta(\t x_1-\t
x_2)\delta(t)
\label{EqG}
\end{eqnarray}
for $\t x_1, \t x_2 > 0$.

The boundary condition $B^T \t{\v \phi}=0$ can be rewritten for $B$ in
the form (\ref{simB}) as
\begin{equation}
\t \phi_a(0^+,t) = -\sum_{b=1}^k (T_0)_{ab}\t \phi_{b+k}(0^+,t),\ \ a=1,
\ldots,k.
\end{equation}
It simply connects the right moving fields $\t \phi_a$ to the left moving
fields $\t \phi_{a+k}$, $a=1,..,k$, through an orthogonal matrix $T_0$.
With this understanding, we find that $G_{ab}$ is given by
\begin{eqnarray}
&& G_{ab}(\t x_1,\t x_2, t)
=\vev{\t \phi_a(\t x_1, t) \t \phi_b(\t x_2,0)}  \nonumber\\
&=& \Big(G_{0}(\t x_1-\t x_2,t) - G_{0}(\t x_1+\t x_2,t) T  \Big)_{ab}
\label{GG0Tt}
\end{eqnarray}
where $T =\bpm{0&T_0\cr T_0^T&0}$.

Note that by restricting to boundary conditions described in terms of
the real bosonic fields, we are ignoring possible symmetries not
present at the Abelian $K$-matrix level.  For instance, if there are
two incoming $\nu = 1$ edges described by Fermi fields $\psi_i$, there
could be a unitary $U(2)$ rotation at $x=0$ rather than the orthogonal
rotation described above.  As an example of the meaning of the orthogonal
matrix $T_0$, consider the case of tunneling between $\nu = 1/3$ and
$\nu = 1$ states discussed in section IV.  The rescaled
basis is $K = {\rm diag}(1,1,-1,-1)$, and the tunneling operator is $\exp(i
(\phi_1 - \sqrt{3} \phi_2)) + {\rm h.c.}$.  The matrix $T_0$ for strong
tunneling is
\begin{equation}
T_0 = \left(\matrix{\frac{1}{2} & \frac{\sqrt{3}}{2} \cr
\frac{\sqrt{3}}{2} & -\frac{1}{2}}\right),
\end{equation}
which is the same as the matrix mapping incoming quasiparticles to
outgoing quasiparticles in Sandler \etal~\cite{sandler}.  For
instance, two incoming electrons on the $\nu = 1$ edge become one
outgoing electron on the $\nu = 1$ edge and three charge $e/3$
quasiparticles on the $\nu = 1/3$ edge.


From the symmetry of the equation for $G_0$,
we see that, as a function of $\t x$,
the matrix function $G_0(\t x, t)$ satisfies
\begin{equation}
 TG_0(\t x,t)T = G_0(-\t x,t)
\label{TGT}
\end{equation}
Using (\ref{TGT}) and $B^T T=B^T$, we can check that the above $G$ satisfies
(\ref{BGGB}). Certainly, $G$ also satisfies the equation (\ref{EqG}).

To obtain the correlation function in the original basis and
in terms of the original labeling $(a, x)$, we need to
start with the explicit form of $G_0$:
$G_0(\t x, t) = -\ln( \Sigma_3 \t x- t)$.
After replacing the label $(a, \t x) $ by $(a, x/|v_a|)$,
we find (no summation over repeated indices)
\begin{eqnarray}
&& G_{ab}(x_1, x_2,t)=\vev{\t \phi_a(x_1, t) \t \phi_b(x_2,0)}  \nonumber\\
&=&  -\ln( (\Sigma_3)_{aa} (x_1/|v_a|- x_2/|v_b|)- t) \delta_{ab} \nonumber\\
&& +
\ln ( (\Sigma_3)_{aa} (x_1/|v_a|+ x_2/|v_b|)- t) (T^T)_{ab}.
\label{GlnT1}
\end{eqnarray}

If $\t K$, $\t V$ and the boundary condition $B$
in the original basis are given by
\begin{eqnarray}
&&\t K = W^T\Sigma_3 W,\ \
\t V = W^T\t V_{diag} W, \cr
&&B \sim W^T \bpm{I\cr T_0^T}
\label{KVB}
\end{eqnarray}
then the correlation function in the original basis can be obtained from the
transformation $\t{\v \phi} \to W^{-1} \t{\v \phi}$,
$\t K =\Sigma_3 \to (W^T) \Sigma_3 W$,
$\t V =\t V_{diag} \to (W^T) \t V_{diag} W$,
$B= \bpm{I\cr T_0^T} \to W^T \bpm{I\cr T_0^T}$, and
$G \to W^{-1}G(W^T)^{-1}$:
\begin{eqnarray}
&&G_{ab}( x_1,  x_2,t) = \nonumber\\
&& - \sum_{cd} (W^{-1})_{ac} \ln( (\Sigma_3)_{cc} (\frac{x_1}{|v_c|}
- \frac{x_2}{|v_d|})-
t)\delta_{cd} (W^T)^{-1}_{db} \nonumber\\
&& + \sum_{cd} (W^{-1})_{ac} \ln ((\Sigma_3)_{cc}(\frac{x_1}{|v_c|}
+ \frac{x_2}{|v_d|})- t) T_{cd} (W^T)^{-1}_{db}.
\label{GlnT}
\end{eqnarray}

The above result for the correlations of $\t {\v \phi}$ allows us to
calculate the correlation functions of vertex operators $O_{\v l}$ by
exponentiation.  Consider an operator
$V_{\v n} = e^{i n^a \t \phi_a}$. Far away from the boundary
($x \gg |v| t$), the operator has a
correlation which is determined by $G_0$ only:
\begin{equation}
\vev{V_{\v n}(x,t) V_{\v n}(x,0)} \sim 1/t^{g_{\v n}}
\end{equation}
where
\begin{equation}
 g_{\v n} = \v n^T \t\Delta \v n
\end{equation}
and $\t\Delta = W^{-1}(W^T)^{-1}$.
We can write $W$ in a form
\begin{equation}
 W = \bpm{\t R_R & 0\cr 0&\t R_L} \t B_{st} \t K_{1/2},
\label{WKRB}
\end{equation}
where $\t R_{R,L} \in O(k)$, and $\t B_{st}$ is the boost matrix of form
\begin{equation}
 \t B_{st} = \exp
\bpm{0 & \t b \cr
\t b^T& 0}.
\end{equation}
In this parameterization of $W$,
$\t\Delta$ depends only on $\t B_{st}$ (or the $k$-dimensional matrix $b$):
\begin{equation}
\t\Delta = \t K_{1/2}^{-1} \t B_{st}^{-2} (\t K_{1/2}^T)^{-1}.
\end{equation}

Near the boundary ($x \ll |v| t$), the
correlation has a different algebraic decay
\begin{equation}
\vev{V_{\v n}(0,t) V_{\v n}(0,0)} \sim 1/t^{g^b_{\v n}}
\end{equation}
with
\begin{equation}
 g^b_{\v n}
=  \v n^T  \t \Delta^b \v n
\end{equation}
where from equation (\ref{GlnT})
\begin{equation}
\t\Delta^b =    W^{-1} (I-T) (W^T)^{-1}.
\end{equation}
The above can be rewritten as
\begin{eqnarray}
\t\Delta^b &=&  2  K^{-1}B (B^T \t\Delta B)^{-1} B^T K^{-1}
\label{DbKB}
\end{eqnarray}
Since it is invariant under $B \to  B U$, (\ref{DbKB}) is valid for all
$B$, not just the form $B=W^T \bpm{I\cr T_0^T}$.

\subsection{Boundary conditions compatible with periodicity conditions}

In the above discussion of the boundary condition (\ref{Bphi}),
we have not considered the problem that this condition violates
the periodicity of the fields $\t {\v\phi}$.
We need to take into account the
periodic nature of $\t{\v \phi}$ field:
\begin{equation}
 \t{\v \phi} \sim \t{\v \phi}^\prime =  \t{\v \phi} + 2\pi \t K^{-1} \v n_c.
\end{equation}
It is clear that the boundary condition
$B^T \t{\v \phi} =0$ is not consistent with all the periodic conditions of
the
$\t{\v \phi}$ field. Since $\t{\v \phi} $ and
$\t{\v \phi} + 2\pi \t K^{-1} \v n_c$ are equivalent, if $B^T \t{\v \phi}
=0$
is allowed, then $B^T (\t{\v \phi} + 2\pi \t K^{-1} \v n_c)  =0$ should also
be allowed. That is, we need to generalize the boundary condition to at least
\begin{equation}
 B^T \t{\v \phi} = 2\pi B^T K^{-1} \v n_c, \ \ \v n_c \in \t \Gamma_c.
\label{GBdry1}
\end{equation}

One technical way to understand what has been done in the previous
section is that we have only considered boundary conditions for the
neutral excitations created by $\partial_x\t {\v \phi}$. (Here
``neutral'' does not mean electrically neutral, but rather conserving
the zero mode of the bosonic theory.)  In addition to these neutral
excitations, there are also charged excitations created by vertex
operators $V_{\v n} = e^{i \v n_c \cdot \t{\v \phi}}$, where $\v n_c$
is a vector in the E-lattice $\t \Gamma_c$.  Since the vertex
operators are the primary fields of the theory, it is the vertex
operators which we expect to have scale-invariant boundary conditions,
rather than the bosonic fields $\t{\v \phi}$.
The generalized boundary condition (\ref{genB}) can also be written as
\begin{equation}
e^{i \v n^T \t {\v \phi}} =1,\ \ \
\v n \in \Gamma_B,
\label{primfields}
\end{equation}
where the rows of $B$ are basis vectors of $\Gamma_B$, or
\begin{equation}
 \Gamma_B =\Latt(B).
\end{equation}
Strictly speaking, it is the boundary condition of the {\it normal-ordered}
exponential which is conformally invariant, and the normal-ordered version
of (\ref{primfields}) has $\infty$ rather than 1 on the right-hand side.

To gain a better understanding of the generalized boundary condition
(\ref{GBdry1}), let us consider a physical realization of the termination of
the edge. We start with an edge described by $\t K$ on $(-\infty, \infty)$.
We then add the following potential term on $(-\infty, 0)$:
\begin{equation}
- \sum_{\v n \in \Gamma_B}
C_{\v n} \cos(\v n\cdot \t {\v \phi})
\label{TermPot}
\end{equation}
where the $k$-dimensional lattice $\Gamma_B$ is a sublattice of $\t \Gamma_c$.
The vectors in $\Gamma_B$ satisfy
\begin{equation}
 \v n^T \t K^{-1} \v n' =0, \ \ \v n,\v n' \in \Gamma_B
\end{equation}
and $C_{\v n}>0$ are very large, so that the potential consistently
pins $\t {\v \phi}$ to the
potential minima, and opens an energy gap in the region $(-\infty, 0)$.
Such a potential leads to the boundary condition
\begin{equation}
 B^T \t {\v \phi} = 2\pi \v n, \ \ \v n \in \Latt(I_{k\times k})
\label{genB}
\end{equation}

{}From the above discussions, we can draw two conclusions. First
not all $B$ matrices are consistent with the periodicity properties of $\t{\v
\phi}$. To specify a valid termination of an edge, $B$ not only must satisfy
(\ref{BKB}), the rows of $B$ must also be in the $\t \Gamma_c$ lattice, or
\begin{equation}
 \Latt(B) \subset \t \Gamma_c
\label{BGac}
\end{equation}
Second, two $B$ matrices, $B_1$ and $B_2$, give rise to the same {\em
generalized} boundary condition if
\begin{equation}
 B_1 =  B_2 M,\ \ M\in GL(k,Z)
\label{Beq}
\end{equation}
Such a pair of $B$ matrices are regarded as equivalent:
\begin{equation}
B_1 \cong B_2
\end{equation}
Note that the above equivalence relation for generalized boundary condition
is stronger (i.e., has smaller equivalence classes)
than the equivalence relation
$B_1 \sim B_2$ defined in (\ref{BsimB}) for the simple boundary condition
$B^T\t{\v \phi}=0$.
{\em The equivalence classes (defined by (\ref{Beq}))
of the $B$ matrices that satisfy (\ref{BKB}) and
(\ref{BGac}) label different terminations (or fixed points) of the edge.}

Now the question is what are the allowed vertex operators on the
boundary.  A boundary vertex operator has the form $V^b_{\v l}=e^{i\v
l^T\t{\v \phi}}$, where $\v l$ is vector in a $k$ dimensional lattice
$\Gamma_{q_B}$ (called the boundary quasiparticle lattice). To determine
$\t \Gamma_{q_B}$ we note that $V^b_{\v l}$ changes one boundary
condition $B^T \t {\v \phi}=2\pi \v n$ to another $B^T=2\pi (\v n +
B^T\t K^{-1} \v l)$.  Thus in order for $ B^T \t {\v \phi}=2\pi (\v n
+ B^T\t K^{-1} \v l)$ to be an allowed boundary condition, $B^T\t
K^{-1} \v l$ must an integral vector. Also, we require that
$V^b_{\v l}$ only shift the combination $ B^T \t {\v \phi}$. In particular,
$V^b_{\v l}$ does not shift the combination $\v l^{\prime T} \t{\v
\phi}$, $\v l' \in \t \Gamma_{q_B}$.  This leads to the condition $\v
l^T \t K^{-1} \v l' =0$ for any $\v l$, $\v l'$ in $\Gamma_{q_B}$.
The above two conditions allows us to determine $\Gamma_{q_B}$:
\begin{eqnarray}
&& \Gamma_{q_B} = \\
&&\Latt(\t K B(B^TB)^{-1} -\frac12 B(B^TB)^{-1}B^T\t KB(B^TB)^{-1}).
\nonumber
\end{eqnarray}
The scaling dimension of $V^b_{\v l}$ is given from (\ref{DbKB}) by
\begin{equation}
h^b(\v l) =  \v l^T  K^{-1}B (B^T \t\Delta B)^{-1} B^T K^{-1} \v l.
\end{equation}

In the above discussion of termination points, we have ignored any symmetry
properties and the related selection rules. In particular, the boundary
condition characterized by $B$ may not conserve the electric charge.
As a result, a boundary vertex operator may not carry a definite electric
charge.
In order for the termination labeled by $B$ to conserve the electric
charge, we must require the $B$ matrix to satisfy
\begin{equation}
 B^T \t K^{-1} \t {\v q} =0.
\label{BCh1}
\end{equation}
In this case, we find that $B^T \t {\v \phi}$,
the fields that are about to be set to a constant, commute with the electric
charge density operator $\rho_e$.
For charge-conserving termination points, the electric charge of a
boundary vertex operator $V^b_{\v l}$ is found to be
\begin{equation}
 Q=\t {\v q}^T \t K^{-1} \v l.
\end{equation}
The condition (\ref{BCh1}) ensures the vertex operators of form
$e^{\v n_B^T\t {\v \phi}}|_{\v n_B \in \Gamma_B}$ are all neutral, so
that they can be set to one without violating the charge conservation.
For a general $B$, the above vertex operators carry nonzero charges
and setting them to one violates the charge conservation.

For a general boundary condition $B$, the charge is not conserved, but
some other quantities may be conserved. On the edge there are $2k$
conserved currents (at low energies) $j_a = \partial_t{ \t \phi_a}$,
one for each branch.  Near the boundary $k$ combinations of the $2k$
conserved currents remain conserved. These $k$ combinations are
given by $B^T\partial_t {\t{\v \phi}}$.  Thus a boundary operator $V^b{\v
l}$ carries $k$ definite combined charges:
\begin{equation}
 \v Q = B^T \t K^{-1} \v l.
\end{equation}

To determine the stability of a fixed point, we also need to know which
boundary operator $V^b_{\v l}$ can appear in the boundary Hamiltonian.
First let us discuss the corresponding issue along the edge.
Along the edge, the lattice $\t \Gamma_c$ label all the mutually local
operators.
Some carry fermionic statistics, and
thus are not allowed in the edge Hamiltonian.
Only the subset described by $\t \Gamma_{c0}$ can appear in the edge
Hamiltonian.
($\t \Gamma_{c0}$ is formed by all the bosonic operators in $\t \Gamma_c$.)
If the charge is conserved, we further require the operators in
$\t \Gamma_{c0}$ to be neutral.

On the boundary, only a subset
of the boundary operators
can appear in the boundary Hamiltonian.
Since there is no statistics within the $0+1$ dimensional boundary,
we only need to check the conservation of the $k$ combined charges.
The values of the combined charge $Q_a$, $a=1,..,k$ allows us to determine
which boundary operators can appear in the boundary Hamiltonian, as in the
examples of the previous section.

\subsection{Image-charge picture and nonchiral fields}

This section shows how the correlation functions can be calculated
from a simple image-charge picture when the chiral bosonic fields are
unified into nonchiral bosons, as is important for a number of
applications.  In particular, we find the falloff of the expectation
value away from the boundary of a vertex operator $e^{i \Phi}$ pinned
to 1 at the boundary, and how the two-body correlations are affected
by the boundary.  The correlation functions of vertex operators found
in subsection B are essentially quite simple: any correlation
function of a vertex operator can be written as a product of
exponentials of correlation functions of free chiral bosons.  One
subtlety is that after rescaling there may be more terms in these
correlation functions than experimental points in the original
problem, since fields at the same physical point become different
points in the rescaled coordinates.  Some additional structure appears
in the correlations when the chiral fields are combined into nonchiral
fields on the half-line, as in the boundary sine-Gordon model.

In practice it is useful to combine the chiral fields $\phi_i$ on the
whole line into nonchiral bosons $\Phi_i$ defined on the half-line
$x<0$, in cases where the fields for $x>0$ are the same as those for
$x<0$.  As an example, the applicability of the integrable boundary
sine-Gordon model used in~\cite{fendley} to determine tunneling
behavior depends on the mapping to the half-line.  The system on the
half-line can be understood as a classical system on the $x>0$ half of
the $(x,t)$ plane, so that the physics known about such
statistical-mechanical systems with boundaries is applicable.  The
technical motivation is that the theory on the half-line will be
invariant at the fixed points under all the conformal generators which
preserve the line $x=0$.  In what follows we show that the correlation
functions of the nonchiral fields can be understood from an ``image
charge'' picture (similar to electrostatics), and that the tunneling
fixed points can be understood as ``ordinary'' and ``extraordinary''
transitions on the half-plane.

For each chiral boson field on the whole line $\phi_j$, define the
nonchiral field $\Phi_j$ on the half-plane $x<0$ by $\Phi_j(z) =
\phi_j(z) + \phi_j({\bar z})$, where $z = t + i x$.
(We use imaginary time $t$ so that conformal invariance is manifest.)
Note that if $z$ has $x>0$, ${\bar z}$
has $x<0$ and that $\Phi_j$ has both left-moving and
right-moving parts.  The vertex operators $\exp(i \alpha \Phi_j)$ will
have different behavior depending on whether $\phi$ changes sign at
$x=0$.  First, with $\eta_j = \pm 1$ the sign gained by $\phi_j$
across $x=0$, for nonzero $\alpha$
\begin{equation}
\langle e^{i \alpha \Phi_j(0,t)} \rangle =
\cases{0 & $\eta = 1$ \cr
\infty & $\eta = -1$}.
\end{equation}
Here and in the sequel we use the normal-ordered exponential, which has
maximum value $\infty$ rather than 1.  Also, below we will consider
the case where $\Phi_j(0,t)$ is not pinned simply to 0 but to some set of
values.  The profile of the order parameter near the boundary can
be calculated simply:
\begin{eqnarray}
\langle e^{i \alpha \Phi_j(x,t)} \rangle &=&
\langle e^{i \alpha \phi_j(x,t)} e^{i \alpha \phi_j(-x,t)}\rangle \cr
&=& \cases{0 & $\eta = 1$ \cr
(2 x)^{-\alpha^2} & $\eta = -1$}.
\end{eqnarray}

The above is the simplest case of the image-charge idea of
Cardy~\cite{cardy,difrancesco}: a correlation function of $n$
nonchiral fields on the half-plane is expressed as a correlation of
$2n$ chiral fields on the full plane.  For the boundary conditions we
are considering, the full-plane correlation functions are known from
(\ref{GlnT}), so the half-plane correlation functions of $\Phi_j$
can be determined.  The two-body function shows different scaling along
the boundary from that in the bulk: (here index $j$ suppressed)
\begin{eqnarray}
&&\langle e^{i \alpha \Phi(z_1)}e^{- i \beta \Phi(z_2)} \rangle \cr
&=& \langle
e^{i \alpha (\phi(z_1) + \phi({\bar z_1}))} e^{-i \beta (\phi(z_2) +
\phi({\bar z_2}))} \rangle \cr
&=& \cases{\delta(\alpha - \beta) \left({4 x_1 x_2 \over
|z_1-z_2|^2 |z_1-{\bar z_2}|^2}\right)^{\alpha^2} & $\eta = 1$ \cr
{|z_1 - {\bar z_2}|^{2 \alpha \beta} \over (2 x_1)^\alpha
(2 x_2)^\beta |z_1 - z_2|^{2 \alpha \beta}} & $\eta = -1$.}
\end{eqnarray}

For example, in the $\eta = 1$ case, the equal-$x$ correlation falls off
as $(t_1 - t_2)^{-4 \alpha^2}$ for $t_1 - t_2 \gg x$, while
far from the boundary ($t_1 - t_2 \ll x$) the falloff is only as
$(t_1 - t_2)^{-2 \alpha^2}$, i.e., with the bulk scaling dimension.
For $\eta = -1$ the correlation along the boundary
is constant at long distances, with leading correction
$(t_1 - t_2)^{-\alpha^2}.$  The critical theory with $\eta = -1$
corresponds to the ``extraordinary'' transition in statistical mechanics,
where the boundary is ordered (the order parameter $\exp(i \Phi)$ has
nonzero expectation value) but the bulk is not, while the $\eta = 1$ theory
corresponds to the ``ordinary'' transition.

Some aspects of the above picture change when the field $\Phi(0,t)$
is pinned to more than one value, e.g., to the minima $\Phi = 2 \pi r
n$, $n \in {\bf Z}$ of $\cos(\Phi / r)$.  Now there is an additional
average over $\Phi_0 = 0, \pm 2 \pi r, \pm 4 \pi r, \ldots$ in the
correlation functions.  The two-body correlation is unchanged, but
the one-body correlation for $\eta = -1$ is now
\begin{equation}
\langle e^{i \alpha \Phi_j(x,t)} \rangle =
\cases{(2x)^{-\alpha^2} & $\alpha = n/r$, $n \in {\bf Z}$ \cr
0 & otherwise},
\end{equation}
which is natural as only those operators invariant under the symmetry
transformation $\Phi \rightarrow \Phi + 2 \pi r$ can have nonzero
expectation values.

\section{Edges with random hopping}

In the $\chi$LL theory, the edge of a bulk QH state with $n$ condensates
is described by two symmetric $n \times n$ matrices, $K$ and $V$.  The
integer matrix $K$ is determined by the bulk QH state and is the same
for all samples of a given edge.  The positive matrix $V$ contains
non-universal velocities and interaction strengths which are expected
to vary from sample to sample.  In this section, we study the RG flow of
the $V$ matrix in the presence of impurity scattering toward fixed
points~\cite{kfp,kane,moorewen} which describe an equilibrated edge..

The RG calculation is given in some detail in an appendix because
there are several new features not present in similar treatments of
the 2D classical XY model~\cite{kosterlitz,jose} and 2D
melting,~\cite{nelson} as well as 1D disordered quantum
electrons.~\cite{giamarchi} Calculations on 1D quantum disordered
systems differ from those on classical 2D systems in that quenched
disorder is random in space but constant in time, so the two spacetime
dimensions enter asymmetrically.  The chirality of the $\chi$LL is
responsible for the differences between our results and previous
results on disordered electrons in 1D: correlation functions in a
$\chi$LL depend on $x + i v t$ rather than just the magnitude $x^2
+ v^2 t^2$, and the operators of interest can have nonzero
``conformal spin'' (difference of right- and left-moving dimensions).
One of the resulting RG equations disagrees with a result previously
obtained by Kane, Fisher, and Polchinski.~\cite{kfp}
We outline our results before proceeding to the calculation

In a maximally chiral edge, such as IQHE edges or $\nu = 2/5$, whether a
given impurity operator (i.e., type of impurity scattering) is relevant
depends only on $K$, not on $V$.
For IQHE edges and also for the main-sequence chiral FQHE edges
$\nu = 2/5, \nu = 3/7, \ldots$, there are relevant impurity operators
which decouple the charge mode from the neutral mode(s).
The charge mode must decouple and the neutral mode velocities
must equilibrate for the system to flow to
the $U(1) \times SU(n)$ fixed point ($n =\dim K$),
where the impurity scattering can
be ``gauged away.''~\cite{kfp,kane}
The $U(1) \times SU(n)$ symmetry possessed by $K$ for these edges~\cite{read}
is generically broken by $V$, but restored if $V$
flows to a decoupled charge mode (the $U(1)$) and $n-1$ neutral modes with
identical velocities (the $SU(n)$).
To our knowledge it has not previously
been shown that general initial conditions flow toward this fixed point
for chiral edges.  The charge mode velocity is {\it not} required to equal
the neutral mode velocity, since the disorder drives $V$ to be diagonal
in a basis where no disorder operator couples charged and neutral modes.
There remain operators which couple the neutral modes to each other;
thus although to leading order in the disorder strength the neutral mode
velocities do not flow together, it seems clear that the eventual
strong-disorder fixed point
will have equal neutral mode velocities but a possibly different charge mode
velocity.  We find that the fixed point is only stable if
the charge mode has greater velocity than the neutral modes.

The main-sequence {\it nonchiral} FQHE edges $\nu = 2/3, 3/5, \ldots$
have similar $U(1) \times SU(n)$ fixed points.  Impurity scattering now
can be either relevant or irrelevant, depending on $V$, and if it
is irrelevant the system will not flow to the fixed point.
For $\nu = 2/3$ KFP used a
perturbative calculation for weak disorder to find the
basin of attraction
of the fixed point~\cite{kfp}; this calculation is similar to ours,
although we find a slight disagreement (Appendix A).
The flow to the fixed point has a much more pronounced effect
on some observable quantities
than in the chiral case: away from the fixed point, conductance
and tunneling properties are nonuniversal.  The differences between
chiral and nonchiral edges result because scaling dimensions
of vertex operators are independent of $V$ (fixed by
$K$) in chiral edges but depend on $V$ in nonchiral edges.

The edge theory of each daughter state of $\nu = 1$ in the hierarchy is
essentially the same as that of the corresponding daughter state of
$\nu = 1/3,1/5,\ldots$.  Every principal hierarchy state, chiral or not,
with neutral modes parallel to each other has a single solvable
charge-decoupled fixed point in the presence of disorder.
Edges with neutral modes traveling in both directions, such as
$\nu = 5/3$ and $\nu = 5/7$, can have infinitely many fixed points
of several different types.~\cite{moorewen}  The RG shows
how for all of these fixed points the charge mode decouples, while
the neutral modes can reach different equilibria, with consequences for
tunneling experiments.  The fixed points not solvable by the KFP method
have disorder operators which frustrate each other at the fixed point.

The $\chi$LL action in imaginary time for a clean edge of a QH state
characterized by the matrix $K$ contains
$n = {\rm dim}\ K$ bosonic fields $\phi_i$:~\cite{wen1}
\begin{equation}
\label{naction}
S_0 = {1 \over 4 \pi} \int{dx \, dt\, [K_{ij} \partial_x \phi_i \partial_t \phi_j +
V_{ij} \partial_x \phi_i \partial_x \phi_j],}
\end{equation}
where the sum over repeated indices is assumed.
$K$ is a symmetric integer matrix and $V$ a symmetric positive matrix. 
$K$ gives the topological properties of the edge: the types of
quasiparticles and their relative statistics.  $V$, the velocity matrix,
is positive
definite so that the Hamiltonian is bounded below.  The electromagnetic
charges of
quasiparticles are specified by an integer vector ${\bf t}$ and the filling
factor is $\nu = t_i (K^{-1})_{ij} t_j$.

Now a term representing quenched random impurity scattering
is added to the action:
\begin{equation}
\label{action2}
S_1 = \int{dx \, dt \, [\xi(x) e^{i m_j \phi_j} + \xi^*(x)
e^{-i m_j \phi_j}]}
\end{equation}
Here $\xi$ is a complex random variable and
$\langle \langle \xi(x) \xi^*(x^\prime)\rangle\rangle = D \delta(x - 
x^\prime)$, with $D$ the (real) disorder strength.
The integer vector ${\bf m}$ describes how many
of each type of quasiparticle are annihilated or created by the operator
$O_{\bf m} = \exp(i m_j \phi_j)$.  For a real system all charge-neutral
scattering operators $m_j$ are expected to appear, but most of these will be
irrelevant in the RG sense.  The condition for charge-neutrality is
$t_i (K^{-1})_{ij} m_j = 0$.
The random variables $\xi_{\bf m}$ for different scattering
operators $O_{\bf m}$ may be
uncorrelated or correlated depending on the nature of the physical
impurities causing the scattering.

The clean action (\ref{naction}) is quadratic and hence does not flow
under RG transformations.  Adding impurities (\ref{action2}) causes
the $V$ matrix in (\ref{naction}) to flow, and in some cases the flow
is to a new type of strong-disorder fixed point.~\cite{kfp,kane,moorewen}
Here we will quickly review the diagonalization of the clean
action to find the eigenmodes ${\bf a}_i$ and their velocities $v_i$,
and then find the RG flows for two-mode edges with a single impurity
operator.  Then a general edge with several modes and impurity operators
is considered.

Let $M_1$ be some matrix which brings $K$ to the pseudo-identity
$I_{n^+,n^-}$: $K^{-1} = M_1 I_{n_+,n_-}
{M_1}^T$.  Then $V$ can be brought to a diagonal matrix
$V_D = {M_2}^T {M_1}^T V M_1 M_2$, where $M_2$ is an element
of the group $SO(n^+,n^-)$ so that $M_2 M_1$ still takes $K$ to the
pseudo-identity.  The point of these transformations is that the action
is now diagonal in the basis ${\bf \tilde\phi} =
(M_1 M_2)^{-1} {\bf \phi}$, so the correlation functions are simple:
\begin{eqnarray}
\langle e^{i \tilde \phi_j(x,t)} e^{- i \tilde \phi_j(0,0)} \rangle &=&
e^{\langle \tilde \phi_j(x,t) \tilde \phi_j(0,0) \rangle - \langle \tilde \phi_j(0,0) \tilde
\phi_j(0,0)\rangle} \nonumber \\
&\propto&
(x \pm i v_j t)^{-1}
\end{eqnarray} 
where the sign depends on whether $\tilde \phi_j$ appears with $-1$ or $+1$ in $I(n^{+},n^{-})$.
The vertex operator $O_{\bf m}$
described by the integer vector ${\bf m}$ has correlation function
\begin{equation}
\langle e^{i m_j \phi_j(x,t)} e^{-i m_j \phi_j(0,0)} \rangle
= \prod^n_{j=1} (x \pm i v_j t)^{-{c_j}^2}
\end{equation}
with $m_j \phi_j = c_j \tilde \phi_j$.  The total scaling dimension
of $O_{\bf m}$ is $\Delta({\bf m}) = \sum c_j^2 / 2$, which is bounded
below by $K({\bf m}) / 2 \equiv {\bf m} K^{-1} {\bf m} / 2$.  The
impurity term $S_1$ containing $O_{\bf m}$ with a random coefficient
 is relevant if $\Delta({\bf m}) < 3/2$; the corresponding
marginal value for a uniform coefficient is $2$, and for a $\delta$-function
coefficient $1$.

Appendix A calculates the change in the correlation function of
$O_{\bf m}$ under an
infinitesimal RG transformation induced by the impurity term $S_1$.
Here we find the change in the underlying $V$ matrix required to
produce the new correlation function.
The $K$ matrix is unchanged as it is ``topological'' (it does not
enter the Hamiltonian).
The $V$ matrix flow has a simple interpretation, valid for any number
of edge modes traveling in either direction.  Each impurity operator
${\bf m}$ drives $V$ to become diagonal in the basis with ${\bf m}$ an
eigenvector.  This automatically minimizes the scaling dimension of
$O_{\bf m}$ in a nonchiral edge.  In cases where there are more impurity
operators than independent eigenvectors, so that not all impurity operators
can simulatenously be eigenvectors, the impurity operators frustrate
each other.

\begin{figure}
\vbox{\epsfxsize=3.25truein
\centerline{\epsffile{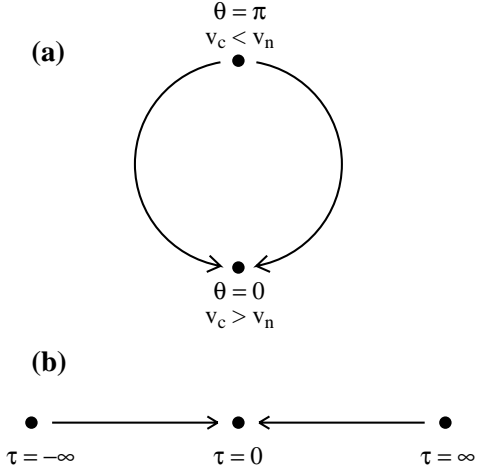}}
\caption{Schematic RG flows for two-mode (a) chiral and (b) nonchiral
edges.  The same idea applies to main-sequence edges with more than two
modes: the chiral case has one stable ($v_c > v_n$)
and one unstable ($v_c < v_n$) fixed point, while
the nonchiral case has one stable fixed point independent of $v_c, v_n$.}
\label{figsix}}
\end{figure}

Both $\nu = 2$ and $\nu = 2/5$ have a single $K({\bf m}) = 2$ operator
which is always relevant.  In the basis ${\bf e}_1 = {\bf t}$, ${\bf
e}_2 = {\bf m},$
\begin{eqnarray}
&&K^{-1} = \left( \matrix{\nu & 0 \cr 0 & 2} \right), \qquad
K^{1/2} = \left( \matrix{1/\sqrt{\nu} & 0 \cr 0 & 1 / 2} \right) \nonumber\\
&&V = K^{1/2} R \left( \matrix{v_1 & 0 \cr 0 & v_2} \right) R^{-1} K^{1/2}.
\label{comovers}
\end{eqnarray}
with $R$ a two-by-two rotation matrix by some angle $\theta$.
Note that all $V$ are obtained by considering $\theta$ in the interval
$[0,\pi)$.
Now $\alpha = 2 \sin^2 \theta$ and $\beta = 2 \cos^2 \theta$
are the exponents appearing in the correlation
function of the impurity operator:
$\langle O_{\bf m}(x,t) O_{\bf m}^\dagger(0,0)\rangle
= (x + i v_1 t)^{-\alpha} (x + i v_2 t)^{-\beta},$
$\alpha + \beta = 2$.
Then to first order in disorder strength, the diagonal velocities
$v_1$ and $v_2$ are unchanged, and (Appendix A)
\begin{equation}
{d\alpha \over d\ell} = - {8 \pi D \alpha \beta \over (v_1 - v_2)
{v_1}^{\alpha - 1} {v_2}^{\beta - 1}}.
\end{equation}
Since $d \alpha = 2 \sin(2 \theta) d\theta = 2 \sqrt{\alpha \beta} d\theta,$
\begin{equation}
{d \theta \over d\ell} = - {4 \pi D \sin(2 \theta) v_1 v_2 \over
(v_1 - v_2) {v_1}^{2 \sin^2 \theta} {v_2}^{2 \cos^2 \theta}}.
\label{thetaflow}
\end{equation}
There are two fixed points of this equation, with $\theta = 0$ stable
and $\theta = \pi / 2$ unstable for $v_1 > v_2$, and vice versa
for $v_1 < v_2$.  The stable fixed point always corresponds to neutral mode
velocity less than charge velocity (Fig.~\ref{figsix}a).

We can summarize the effect of the disorder operator in the comoving case
simply: it rotates $V$ so that ${\bf m}$ becomes
an eigenvector.  Since ${\bf m}$ is neutral (${\bf m}K^{-1} {\bf t}$ = 0)
the other eigenvector is driven to the charge vector.  The idea that
impurity operators drive $V$ to make themselves eigenvectors is quite general.
The case of two {\it countermoving} modes (e.g., $\nu = 2/3$) with a
$K({\bf m}) = -2$ disorder operator is similar in form.
The rotation matrix $R$ in (\ref{comovers}) is replaced
by a boost matrix $B$,
\begin{equation}
B = \left( \matrix{\cosh \tau & \sinh \tau \cr \sinh \tau & \cosh \tau}
\right),
\end{equation}
and the exponents in the correlator are $\alpha = 2 \sinh^2 \tau$,
$\beta = 2 \cosh^2 \tau,$ $\alpha - \beta = -2$, $\alpha + \beta =
2 \Delta({\bf m}).$  The flow equation for $\tau$ is then
\begin{equation}
{d\tau \over d\ell} = - {4 \pi D \sinh(2 \tau) v_+ v_- \over (v_+ + v_-)
v_+^{2 \sinh^2 \tau} v_-^{2 \cosh^2 \tau}}.
\end{equation}
Here $v_+$ and $v_-$ are the (positive) velocities of the
right- and left-moving modes.  Now there is only one fixed point,
at $\tau = 0$ (Fig.~\ref{figsix}b), which is the solvable fixed point
found by KFP.~\cite{kfp}

Now we consider a general edge with several modes and impurity operators.
To first order in disorder strength, the effects of each impurity operator
add independently.
Scaling dimensions $\Delta({\bf m})$ of vertex operators $O_{\bf m}$
are independent of $V$ in chiral
edges, depending only on $K$: $2 \Delta({\bf m}) =
{\bf m} K^{-1} {\bf m} = K({\bf m})$.
In a nonchiral edge this holds as an inequality: $2 \Delta({\bf m}) \geq
K({\bf m})$, with equality only if $V$ is diagonal in a
basis with ${\bf m}$ an eigenvector.  Since most experimentally
relevant quantities are determined by scaling dimensions, it is useful
to isolate which parts of $V$ affect scaling dimensions.
The matrix $M_2$ used above to diagonalize $V$ while preserving the
pseudo-identity $I_{n^+,n^-}$ is an element of $SO(n^+,n^-)$.  It can be
decomposed $M_2 = B R$ into a product of a `boost', a symmetric matrix, and a
rotation, an orthogonal matrix, each elements of
$SO(n^+,n^-)$.~\cite{moorewen}
The $n(n+1)/2$ free parameters in the symmetric positive matrix $V$ are now
taken as $n$ eigenvelocities (the elements of $V_D$), $n^+ n^-$ ``boost
parameters'' (which correspond to interactions between oppositely
directed modes), $n^+ (n^+ - 1) / 2$ rotation parameters between
right-movers, and $n^- (n^- - 1) / 2$ rotation parameters between
left-movers.  Only the $n^+ n^-$ boost parameters affect scaling dimensions
because, introducing the matrix $\Delta_{ij}$ via
$2 \Delta({\bf m}) = {\bf m}_i \Delta_{ij} {\bf m}_j$,
\begin{equation}
\Delta = M_1 B R R^T B^T
{M_1}^T = M_1 B^2 {M_1}^T.
\end{equation}
For each pair of comoving modes appearing in the correlation function,
there is an infinitesimal change in the rotational part $R$, and
for each pair of countermoving modes, there is an infinitesimal change in
the boost part $B$.

Given an impurity operator $O_{\bf m}$ and initial $V(\ell)$, we need to find
$V(\ell + d\ell)$ which gives the changes to the correlation function
calculated in Appendix A.  However, there is an important issue
not present in the two-mode case: there are more free parameters in $V$
than exponents in the correlation function, so $V$ is not uniquely
determined without additional assumptions.  We assume that
each term in Appendix A coupling two modes affects only the
components of $V$ between those two modes.  We will also write the flow
equations for the components of $V$ directly, rather than introducing
a parametrization as we did above in terms of $\theta$ or $\tau$,
because for multi-mode edges such parametrizations become quite complicated.

Suppose that under an infinitesimal change $d\ell$
the $V$ matrix changes from $(M_1 M_2)^{-1 T}
V_D (M_1 M_2)^{-1}$ to
\begin{equation}
V(\ell + d\ell) =
(M_1 M_2 N)^{-1 T}
V_D (M_1 M_2 N)^{-1}.
\label{ndef}
\end{equation}
Note that $V$ does not flow if the eigenmodes have the same velocity,
so that $V_D$ is a multiple of the identity.
Here $N$ is an element of $SO(n_+,n_-)$ differing only by order $d\ell$
from the identity.  The fields which diagonalize $V(\ell + d\ell)$ are
$\phi^\prime = (M_1 M_2 N)^{-1} \phi = N^{-1} {\tilde \phi}$.
Now let $c_i$ be the components of the disorder operator in the
${\tilde \phi}$ basis which diagonalizes $V(\ell)$:
$m_i \phi_i = c_j {\tilde \phi_j}$.  The exponents appearing in the
correlation function, whose flow is calculated in Appendix A, are
$\alpha = {c_1}^2$, $\beta = {c_2}^2$, etc.

Now we are ready to construct $N$.
For each pair of countermoving modes $i$ and $j$,
$N$ has an infinitesimal rotation angle $d\theta_{ij} = -d\theta_{ji}$,
and for each pair of comoving modes $i$ and $j$, a boost
$d\tau_{ij} = d\tau_{ji}$.  For an edge with two right-movers and one
left-mover, e.g.,
\begin{equation}
N = \left( \matrix{1&-d\theta_{12} & d\tau_{13}\cr
d\theta_{12}&1&d\tau_{23}\cr
d\tau_{13}&d\tau_{23}&1} \right).
\end{equation}
Then for a disorder operator $O_{\bf m}$ with projections $c_i$
and strength $D$,
\begin{eqnarray}
d\theta_{ij} &=& - {4 \pi c_i c_j v_i v_j
D \, d\ell\over (v_i - v_j) \prod_i {v_i}^{{c_i}^2}}; \nonumber\\
d\tau_{ij} &=& - {4 \pi c_i c_j v_i v_j D \, d\ell \over
(v_i + v_j) \prod_i {v_i}^{{c_i}^2}}.
\end{eqnarray}
Note that the $c_i$ can have either sign and must change sign in the
vicinity of a fixed point, as above in the two-mode case.  Substituting
$N$ in (\ref{ndef})
gives the desired RG flow equation for the components of $V$.
The condition for a fixed point is simple: $N^T V_D N = V_D$.
Hence for any chiral edge ($N$ a pure rotation), if all velocities are
the same the system is at a fixed point.

The simplest multicomponent edges are the IQHE edges $\nu = n$ and
chiral main-sequence edges $\nu = n/(2n+1)$.  The behavior
of these is similar to the $\nu=2$ case discussed in detail above: the
stable fixed point has charge and neutral mode velocities $v_c > v_n$.
The neutral mode velocities are expected to equalize while the charge
mode remains different, because there are always hopping operators connecting
the different neutral modes, while there is no hopping connecting the charge
mode to the neutral modes.  As a result there is no stable
fixed point unless the neutral velocities are the same.

For the $\nu = 3$
case we can demonstrate the similarity to the $\nu = 2$ case by an explicit
calculation.  Parametrize the rotation part $R$ of $V$ as
$R = R_{23}(\theta_1) R_{12}(\theta_2) R_{23}(\theta_3)$, where
$R_{ij}(\theta)$ is the rotation by $\theta$ in the $i-j$ plane; then in
a basis with ${\bf e}_1 = {\bf t}$, the charge mode is decoupled if
$\theta_2 = 0$.  The flow equation for $\theta_2$, with $v_2 = v_3 = v_n$,
is
\begin{eqnarray}
&&{d\theta_2 \over d\ell} =
{-4 \pi v_c v_n \sin(2 \theta_2) \over (v_c - v_n)} \times \nonumber\\
&&\Bigg[ {\cos^2 \theta_1 (4 D_1 + D_2 + D_3) \over 4
{v_c}^{2 \cos^2 \theta_1 \sin^2 \theta_2} {v_n}^{2 (\sin^2 \theta_1
+ \cos^2 \theta_1 \cos^2 \theta_2)}} \nonumber \\
&&+ {3 \sin^2 \theta_1 (D_2 + D_3) \over
4 {v_c}^{2 \sin^2 \theta_1 \sin^2 \theta_2} {v_n}^{2 (\cos^2 \theta_1
+ \sin^2 \theta_1 \cos^2 \theta_2)}} \Bigg]
\end{eqnarray}
where $D_{1\ldots3}$ are the strengths of the 3 hopping operators.
This is of the same form as (\ref{thetaflow}) since the quantity in
brackets is clearly positive.  Henceforth we will not write out
the flow equations but just discuss the qualitative behavior.

The nonchiral main-sequence edges $\nu = 2/3$, $3/5$,$\ldots$ have
solvable stable $U(1) \times SU(n)$ fixed points if all the neutral mode
velocities are equal and the charge mode is decoupled.
The charge mode does indeed decouple in the perturbative RG equations
since each impurity operator reduces the scaling dimension of ${\bf t}$
toward its minimum $\nu$, which is only attained when the charge mode
is decoupled.  Once the charge mode is decoupled, the neutral modes
behave exactly as in the chiral case, except that
the fixed point is stable even if the neutral mode velocity is greater
than the charge mode velocity.

In the edges $\nu = 5/3$ or $\nu = 5/7$,
which have neutral modes in both directions, there are an infinite number
of possibly relevant impurity operators.~\cite{moorewen}  However, near each of
the possible fixed points there are only three relevant operators.  Unlike
the case with all neutral modes in the same direction,
to first order in disorder strength there is no stable
fixed point for $V$, even if the neutral mode velocities are equal.
The mathematical difference is that, now that the modes move in opposite
directions, the impurity operators cause infinitesimal boosts which
do not disappear when the velocities are equal, unlike infinitesimal
rotations.  The fixed points in these edges have marginal
operators not present for the KFP-type fixed points
and seem to be of a different
type, and their strong-disorder behavior and stability is
not well understood.

There are several four-component edges, such as $\nu = 12/17$ and
$\nu = 12/31$, which have solvable, $SU(n)$ symmetric fixed points of
the KFP type as well as fixed points with marginal operators similar
to the $\nu = 5/3$ and $\nu = 5/7$ edges.
The picture of equilibration differs depending on the fixed point. 
The solvable $SU(2) \times SU(2)$ fixed points found in these edges can
have all three neutral mode velocities different (since there is a basis
where no relevant hopping operator couples neutral modes),
while the $SU(3)$ fixed points have to have two neutral mode velocities
equal in order to be solvable.
However, as discussed
in the introduction the fixed point accessible by the composite-fermion
approach~\cite{shytov} is the one with marginal operators present,
even though this is the only one of the three fixed points whose stability
is doubtful.~\cite{moorewen}

\section{conclusions}

The main result of this paper is the analysis for a general Abelian
edge of a large class of tunneling/termination fixed points.  We find
the conditions on allowable boundary conditions and the resulting
correlation functions for vertex operators at the fixed points.  We
find that the requirement of unitary time evolution imposes conditions
on when a quantum Hall edge can terminate at a point or join smoothly
to another edge.  Two edges can join smoothly only if they have the
same number of modes, and there are additional statistical
restrictions: for example, $\nu = 1$ can join smoothly to $\nu = 1/9$
(provided a source of charge is present), but not to $\nu = 1/3$.  We
find the lattice of allowed boundary operators, both charged and
neutral, for a given boundary condition.

The correlation functions in the presence of a boundary can be
understood through an ``image-charge'' picture, and take a simple form
which becomes more complex when the original chiral bosons are
combined into nonchiral fields as for the boundary sine-Gordon model.
The two-point correlation function of a vertex operator can show different
behavior (a change in scaling dimension) as the two locations
are moved toward the boundary.

We solve a simplified model of how interactions between electrons on
different edges affect tunneling through a point junction.  Most
current tunneling experiments involve multiple contacts between the
two edges and variable interaction strengths along the edge, but
the basic result that interactions can complicate the
identification of the effective tunneling charge should still apply.
The model we consider maps onto the solvable model of Fendley \etal with
a {\it continuous} effective filling fraction, giving a physical
realization of this result beyond the discrete cases $\nu_{{\rm eff}}
= 1$,$1/2$,$1/3$ known previously.

We have used a perturbative RG approach to find the flows of random
hopping operators in a general quantum Hall edge.  Two broad
conclusions from this treatment are that with no initial assumptions
favoring one type of state over another, significant differences
emerge between non-principal states and principal states, and between
states with all neutral modes in one direction and states with neutral
modes in both directions.  In states with neutral modes in both
directions, different hopping operators compete to drive the system to
different fixed points.  This may explain why hierarchy quantum Hall
states with neutral modes in both directions are much more difficult
to observe than those at the same level of the hierarchy with all
neutral modes in one direction, and why non-principal states are also
rarely seen.  We find that charge-neutral separation is a general feature
when disorder is relevant, in both chiral and nonchiral edges.  A
specific result for the $\nu = n$ IQHE states and chiral main-sequence
FQHE states is that the charge-neutral separated fixed point is only
perturbatively stable when the charge mode is faster than the neutral
mode, as for strong Coulomb interactions.

X.G.W acknowledges support by NSF under the MRSEL
Program DMR 98-08491 and by NSF grant No. DMR 97-14198.

\appendix
\section{Perturbative Renormalization Group}

This appendix uses the perturbative RG technique to study the effect of
impurity scattering operators
on the velocity matrix $V$ in the chiral-Luttinger-liquid action
(\ref{action}).  The $K$ matrix does not flow and thus remains an integer
matrix, which follows directly from the fact that $K$ does not enter the
Hamiltonian (i.e., it is ``purely topological'').  The constancy of
$K$ to leading order in $D$ will be explicit
in the results obtained below.  In order to calculate the
changes in $V$ under a change in the cutoff, we expand a correlation
function to first order in the disorder strength, then show that the
terms proportional to the disorder strength can be interpreted as
infinitesimal changes in the matrix $V$.

The real-space calculation is similar to previous RG calculations on
2D classical models~\cite{kosterlitz,jose,nelson} and
1D electrons~\cite{giamarchi}.  The differences arise from the chiral
nature of quantum
Hall edge states.  As an example, consider the correlation function
of a vertex operator $O_{\bf m} = \exp(i m_j \phi_j)$ in a nonchiral edge
with one mode in each direction:
\begin{equation}
\langle O_{\bf m}(r_1) O_{\bf m}^\dagger(r_2) \rangle \propto
(x + i v_+ t)^{-\alpha} (x - i v_- t)^{-\beta}
\label{correlation}
\end{equation}
with $\alpha - \beta = K({\bf m})$ an even integer, $r_i = (x_i,t_i)$,
$x = x_2 - x_1$ and
$t = t_2 - t_1$, and $(v_+,v_-)$ the
velocities of the right and left moving modes.  Unless $K({\bf m}) = 0$ and
$v_+ = v_-$, the correlation function has a phase as well as
a magnitude.

First we treat the case of an edge with
two modes, either parallel or antiparallel, and then show how the
flows for an edge with more than two modes follow with no further
computation.

The correlation function of an operator $O_{\bf n}$,
expanded to first order in the disorder strength, is
\begin{eqnarray}
&&\langle e^{i n_j \phi_j(r_1)} e^{-i n_j \phi_j(r_2)} \rangle_1
= \langle e^{i n_j \phi_j(r_1)} e^{-i n_j \phi_j(r_2)} \rangle_0 \times \cr
&&(1 - \int dr_3\,dr_4\,[\xi(x_3) \xi^*(x_4)  
\langle e^{i m_j \phi_j(r_3)} e^{-i m_j \phi_j(r_4)} \rangle_0]) \cr
&&+\int dr_3\,dr_4\,[\xi(x_3) \xi^*(x_4) \times \cr
&&\quad\quad\langle e^{i n_j \phi_j(r_1)} e^{-i n_j \phi_j(r_2)}
e^{i m_j \phi_j(r_3)} e^{-i m_j \phi_j(r_4)} \rangle_0].
\label{starteq}
\end{eqnarray}
Now carry out the disorder average $\{ \xi(x) \xi^*(x^\prime) \}
= D \delta(x - x^\prime)$ and consider the term with four correlation
functions.
Introduce ${\bf R} = ({\bf r}_3 + {\bf r}_4)/2, {\bf r} = {\bf r}_4 -
{\bf r}_3.$  Only configurations where the internal points ${\bf r}_3$
and ${\bf r}_4$ are near each other (i.e.,
separated by the cutoff $a$) contribute to the RG
flows.~\cite{nelson}
At this point assume for convenience that we are calculating
the correlation function of the disorder operator itself
$({\bf m} = {\bf n})$.  The symbol $P_{12}$ denotes
$(x + i v_+ t)^{-\alpha} (x - i v_- t)^{-\beta},
x = x_2 - x_1, t = t_2 - t_1$.
Because $\alpha - \beta = K({\bf m})$ is
even, $P_{12} = P_{21}$.  The last term in (\ref{starteq}) is now
\begin{equation}
D \int dX\,dT\,dt\,[{P_{12} P_{34} P_{14} P_{23} \over
P_{13} P_{24}}].
\end{equation}
The integrand is $P_{12} P_{34}
\exp(g_{13} + g_{24} - g_{14} - g_{23}) \approx P_{12} P_{34}
\exp({\bf r} \cdot \nabla_{\bf R} (g({\bf r}_1 - {\bf R}) -
g({\bf r}_2 - {\bf R}))).$ 
Here
$g_{ab} \equiv -\log(P_{ab}) = \alpha \log(x + i v_+ t) + \beta \log(x - i v_- t).$
Disorder fixes $x = 0$ in ${\bf r} = (x,t)$ so
the exponential is $\exp(t \partial_T (g({\bf r}_1 - {\bf R}) -
g({\bf r}_2 - {\bf R}))) \approx 1 + t^2
(\partial_T g({\bf r}_1 - {\bf R}) - \partial_T g({\bf r}_2 - {\bf R}))^2$.
Hence we need to evaluate the following integral:
\begin{eqnarray}
&&D \int dX\,dT\,dt\,P_{12} P_{34}
[1 + t^2 ({i \alpha v_+ \over X - x_1 + i v_+ (T - t_1)} \cr
&&\quad - {i \beta v_- \over X - x_1 - i v_- (T - t_1)}
- {i \alpha v_+ \over X - x_2 + i v_+ (T - t_2)}\cr
&&\quad\quad+ {i \beta v_- \over X - x_2 - i v_- (T - t_2)})^2].
\label{squares}
\end{eqnarray}
The constant term in the integrand cancels the leading term in the
partition function in (\ref{starteq}).  When the square is expanded,
products of denominators at the same point will cancel infinities
arising in the calculation of products of denominators at
different points, leaving a finite answer.  The change of cutoff in the
$t$ integral will yield the RG equations at the end.

First consider the integrals of the $\alpha^2, \beta^2$ terms.
After rescaling the time variables by $v_+$, the $\alpha^2$
integral is
\begin{eqnarray}
&&\int{dX \, dT \, [{1 \over X + i T - x_1 - i t_1}
\,{1 \over X + i T - x_2 - i t_2}]} \cr
&&= \int{dX \, dT \, [{1 \over z^2 - w^2}]},
\label{badint}
\end{eqnarray}
with $z = X + i T, 2 w = x_2 - x_1 + i t_2 - i t_1.$  But this integral
is not uniformly convergent at $\infty$ and thus not well-defined:
for example, if $w = i$ the
integral is $2 \pi$ if the $X$ integration is done first, $0$ if the
$T$ integration is done first, and $\pi$ if the integration is done
in radial coordinates.  We believe that the appropriate value of the
integral (\ref{badint}) is 0, because in Minkowski space
(real rather than imaginary time) the corresponding
integral has integrand $(x + t - x_1 - t_1)^{-1} (x + t - x_2 - t_2)^{-1}$
and is unambiguously zero.  Also, zero is the only value consistent
with the fact that the randomness can be rotated away at the KFP fixed
point, since at that point the RG flow of the velocity should be
independent of the disorder strength.

The $\alpha \beta$ terms, give the renormalization of the
scaling dimension $\Delta$, are proportional to
\begin{eqnarray}
I&=& \int dX \, dT \, [{1 \over c_- d_+} + {1 \over c_+ d_-}],\cr
&&c_\pm = X - x_1 \pm i v_\pm (T - t_1), \cr
&&d_\pm = x_2 - X \pm i v_\pm (t_2 - T).
\label{goodint}
\end{eqnarray}
First do the $dX$ integral as a contour integral.  The poles of the
first term are at $w_1 = x_1 + i v_- (T - t_1)$ and $w_2 = x_2 + i v_+ (t_2 - T)$, and the integral vanishes unless the poles are on different
sides of the real axis (likewise for the second term).
Thus $T \notin [t_1, t_2]$ and we are left with
\begin{eqnarray}
I &=& \left(\int_{- \infty}^{t_1} dT + \int_\infty^{t_2} dT\right)
[{2 \pi i \over y_1} - {2 \pi i \over y_2}], \cr
y_1 &=& x_2 - x_1 + i v_+ t_2 + i v_- t_1 -
i (v_+ + v_-) T,\cr
y_2 &=& x_2 - x_1 - i v_+ t_1 - i v_- t_2 + i
(v_+ + v_-) T.
\end{eqnarray}
Hence
\begin{eqnarray}
I &=& {2 \pi \over v_+ + v_-} \log \left( {a_1^+ a_1^- a_2^+ a_2^- \over
(b^+ b^-)^2}\right),\cr
&&a_1^\pm = x_2 - x_1 \pm i (v_+ + v_-) \infty + i v_+ t_2 + i v_- t_1\cr
&&a_2^\pm = x_2 - x_1 \pm i (v_+ + v_-) \infty - i v_+ t_1 - i v_- t_2,\cr
&&b^\pm = x_2 - x_1 \pm i v_\pm (t_2 - t_1).
\end{eqnarray}
The finite part of the result is independent of the order of integration,
even though (\ref{goodint}) is superficially even less well-defined than
(\ref{badint}).  The infinite part of the result is canceled by the
$\alpha \beta$ terms in (\ref{squares}) with denominators at the
same point.  The $dt$ integral is
\begin{eqnarray}
\int_{- \infty}^\infty dt\,t^2 P_{34} &=& 2 \int_a^\infty dt\,[t^2
({1 \over i v_+ t})^\alpha
({1 \over - i v_- t})^\beta] \cr
&=& i^{K({\bf m})}
{2 \over {v_+}^\alpha {v_-}^\beta}
\int_a^\infty dt\,t^{2-(\alpha + \beta)}. 
\end{eqnarray}
The dependence on $K({\bf m})$ here is an artifact of our using the
unfortunate convention
(\ref{correlation}), when in fact the proportionality constant in front
alternates sign as $|K({\bf m})| = 0, 2, 4, \ldots$
to keep the correlation function positive when its
argument is on the time axis.
The effective scaling dimension $\Delta_{\rm eff}$ after
re-exponentiating the perturbation to the correlation function is
\begin{equation}
2 \Delta_{\rm eff} = 2 \Delta - {4 (4 \pi) \alpha \beta D \over (v_+ + v_-)
{v_+}^{\alpha - 1} {v_-}^{\beta - 1}} \int_a^\infty {dt \over t^{\alpha + \beta - 2}}.
\end{equation}
By the usual process of changing the cutoff $a \rightarrow a \exp(\ell)$
we obtain the RG equations
\begin{eqnarray}
{dD \over d\ell} &=& (3 - 2 \Delta) D \cr
{d\Delta \over d\ell} &=& - {8 \pi D (\Delta^2 - K({\bf m})^2 / 4) \over
(v_+ + v_-) {v_+}^{\alpha-1} {v_-}^{\beta-1}}.
\label{counterflows}
\end{eqnarray}
These equations match those found by Kane, Fisher, and Polchinski
for the $|K({\bf m})| = 2$ operator in the $\nu = 2/3$ state.  However,
as mentioned above we find no term which renormalizes the velocities to
first order in D.

Now consider the case of two comoving modes.  The simplest example is
the IQHE state $\nu = 2$, which has a relevant impurity operator
hopping electrons from one mode to the other.  The correlation function
$P_{12}$ of an impurity operator $O_{\bf n} = \exp(i n_j \phi_j)$
has the form $(x + i v_1 t)^{-\alpha} (x + i v_2 t)^{-\beta}$,
with $(v_1,v_2)$ the velocities of the two eigenmodes and
$\alpha + \beta = K({\bf n})$ an even integer.  Expanding the correlation
function in the disorder strength and then evaluating
the disorder average as before
gives the correction to the correlation function:
\begin{eqnarray}
&&D \int dX\,dT\,dt\, P_{12} P_{34}
[1 + t^2 ({i \alpha v_1 \over X - x_1 + i v_1 (T - t_1)} \cr
&&\quad+ {i \beta v_2 \over X - x_1 + i v_2 (T - t_1)}
- {i \alpha v_1 \over X - x_2 + i v_1 (T - t_2)} \cr
&&\quad\quad - {i \beta v_2 \over X - x_2 + i v_2 (T - t_2)})^2].
\end{eqnarray}
Expanding the square gives terms with both denominators at the same
point, which cancel infinities appearing elsewhere in the calculation,
and terms with both denominators having the same velocity, which were
previously argued to be zero
(and in any event cannot cause the two velocities to flow toward each other,
since each term only involves one velocity).
The result is
\begin{eqnarray}
&&\alpha \beta v_1 v_2 P_{12}
\int dX\,dT\, [{1 \over X - x_1 + i v_1 (T - t_1)}
\times \cr
&&\quad{1 \over X - x_2 + i v_2 (T - t_2)} + (v_1 \leftrightarrow v_2)]
\int dt\,t^2 P_{34} \cr
&&= {4 \pi \alpha \beta v_1 v_2 P_{12}  \int dt\,t^2 P_{34} \over v_1 - v_2}
\log[{x_2 - x_1 + i v_1 (t_2 - t_1) \over
x_2 - x_1 + i v_2 (t_2 - t_1)}].
\end{eqnarray}
Thus $\alpha$ and $\beta$ are changed
but not $\alpha + \beta = K({\bf m})$.  The velocities of the
eigenmodes are unaltered, and the RG flows are
\begin{eqnarray}
{d D \over d\ell} &=& (3 - \alpha - \beta) D = (3 - K({\bf m})) D \cr 
{d \alpha \over d\ell} &=& - {d \beta \over d\ell} = - {8 \pi D \alpha \beta
\over (v_1 - v_2) {v_1}^{\alpha - 1} {v_2}^{\beta - 1}}.
\label{coflows}
\end{eqnarray}
The singular denominator when $v_1 = v_2$ is acceptable because at
$v_1 = v_2$, only $\alpha + \beta$ is well-defined, not
$\alpha$ and $\beta$ separately.

The generalization to a case with two modes and more than one impurity
operator is simple: the contributions to the RG flow equations for
the velocity matrix from each impurity operator add, since to leading
order the impurity operators are independent.

Extending the calculation to an edge with more than two modes is
quite simple.  In the expansion of the square term in (\ref{squares}),
each pair of modes gives one term.  If the two modes move in opposite
directions, the term lowers the total scaling dimension as in
(\ref{counterflows}); if the two modes move in the same direction,
the term maintains the total scaling dimension as in (\ref{coflows}).
Each term preserves $K({\bf m})$ separately. 
For explicitness, consider the case of an edge with two right-movers and one
left-mover, which is relevant to the $\nu = 3/5$ edge.
Then the correlation function has the form $(x + i v_1 t)^{-\alpha}
(x + i v_2 t)^{-\beta} (x - i v_- t)^{-\gamma}$, and the RG flows for the
exponents are
\begin{eqnarray}
{d \alpha \over d\ell} &=& - {8 \pi D \alpha \beta v_1 v_2 \over (v_1 - v_2)
{v_1}^\alpha {v_2}^\beta {v_-}^\gamma}
- {8 \pi D \alpha \gamma v_1 v_- \over (v_1 + v_-) 
{v_1}^\alpha {v_2}^\beta {v_-}^\gamma} \cr
{d \beta \over d\ell} &=&  {8 \pi D \alpha \beta v_1 v_2 \over (v_1 - v_2)
{v_1}^\alpha {v_2}^\beta {v_-}^\gamma}
- {8 \pi D \beta \gamma v_2 v_- \over (v_2 + v_-)
{v_1}^\alpha {v_2}^\beta {v_-}^\gamma} \cr
{d \gamma \over d\ell} &=& - {8 \pi D \alpha \gamma v_1 v_- \over (v_1 + v_-) 
{v_1}^\alpha {v_2}^\beta {v_-}^\gamma}
- {8 \pi D \beta \gamma v_2 v_- \over (v_2 + v_-)
{v_1}^\alpha {v_2}^\beta {v_-}^\gamma}.
\end{eqnarray}

Section VI shows how the $V$ matrix flow in the
$\chi$LL action determined by the correlation function flows found
in this appendix has a natural interpretation in terms of the
boost and rotation parts of the $V$ matrix.

\end{document}